\documentclass[12pt]{article}
\usepackage{float}
\usepackage{amsfonts}
\usepackage{bm}
\usepackage{amsmath}
\usepackage{amscd}
\usepackage{amssymb,lscape}
\usepackage{latexsym,layout,amsfonts,amssymb,fontenc,xcolor,amsthm,multirow,
amsfonts,bm,textcomp}
\usepackage{caption}

\newcommand{\bmat}{\left(\begin{array}}
\newcommand{\emat}{\end{array}\right)}
\def\gtrsim{\mathrel{\raise.3ex\hbox{$>$\kern-.75em\lower1ex\hbox{$\sim$}}
}
}

\def\p{\partial}

\def\ap{\alpha^{\prime}}

\def\-{\hphantom{-}}

\def\s2{\frac{1}{\sqrt2}}

\def\beq{\begin{equation}}
\def\eeq{\end{equation}}
\def\beqa{\begin{eqnarray}}
\def\eeqa{\end{eqnarray}}

\def\mg{m_{3/2}}
\def\mg2{m^2_{3/2}}

\def\Dsl{\,\raise.15ex\hbox{/}\mkern-13.5mu D} 

\def\be{\begin{equation}}
\def\ee{\end{equation}}
\def\bea{\begin{eqnarray}}
\def\eea{\end{eqnarray}}

\DeclareMathOperator{\Spin}{\mathit{Spin}}

\newcommand{\nn}{\nonumber}

\makeatletter
\@addtoreset{equation}{section}
\makeatother

\begin{document}
\pagestyle{plain}
\begin{titlepage}
\begin{center}
  \LARGE{ Symmetry enhancement interpolation, non-commutativity and Double 
Field 
Theory
\\[6mm]}
\large{\bf  G. Aldazabal${}^{a,b,c}$, E. Andr\'es$^{a,c}$, M.Mayo${}^{a,b}$, V. 
Penas${}^{a,b}$
 \\}
\small{ ${}^a$ {\em G. F\'isica CAB-CNEA, }\\{\em Centro At\'omico 
Bariloche, Av. Bustillo 9500, Bariloche, 
Argentina.}\\ 
${}^b${\em Consejo Nacional de Investigaciones Científicas y Técnicas 
(CONICET)} \\
${}^c${\em Instituto Balseiro,Universidad Nacional de Cuyo (UNCUYO) } \\[-0.3em]
{\em  Av. Bustillo 9500, R8402AGP,  Bariloche, 
Argentina.}\\[-0.3em]}
\end{center}
{\scriptsize{{{E-mail: {aldazaba@cab.cnea.gov.ar}, 
andres@cab.cnea.gov.ar, {martin.mayo@ib.edu.ar}, 
{victor.penas@cab.cnea.gov.ar}}}}}
\vspace{0.5cm}

\noindent
 {\bf Abstract}:
We present a moduli dependent target space 
effective field theory action for (truncated) heterotic string toroidal 
compactifications.  When moving continuously along moduli space,
the stringy gauge symmetry enhancement-breaking effects, which occur at 
particular points of moduli space, are reproduced.

Besides the expected fields, originating in the  ten dimensional low 
energy effective theory, new vector   and scalar fields are 
included.  These fields depend on ``double periodic coordinates'' as usually 
introduced  in Double Field Theory. Their mode expansion encodes information 
about string states, carrying  winding and KK momenta, associated to gauge 
symmetry enhancements. It is found that a non-commutative 
product, which introduces an intrinsic non-commutativity on the 
compact target space, 
is required in order to make contact with string theory amplitude results.

\vspace{.5cm}
\today


\end{titlepage}


\begin{small}
\tableofcontents
\end{small}

\newpage\section{Introduction}
\label{sec:Introduction}

In this article we propose a target space 
effective field theory  description of string theory interactions. 
Clearly the subject is not new. Indeed, the  conventional low energy  
effective  action for given values of moduli fields can be found in string 
books \cite{Green:1987sp,Blumenhagen:2013fgp,iu}. 
However, several works point towards a richer structure with 
some intrinsic compact target space non-commutativity. Among them, there are 
recent analyses \cite{agimnr,aamr,Cagnacci:2017ulc,aamp,Fraiman} performed 
from the perspective of 
Double Field Theory (DFT)\footnote{
See \cite{DFTreviews, Hull:2009mi} for some 
original references on DFT and for instance \cite{reviewamn,reviewOLZ} 
for reviews. DFT approaches to heterotic string can be found in 
\cite{dftheterotic}.}  
 aiming at 
the inclusion of gauge symmetry enhancement aspects in the field theory 
description,  as well as 
recent (and not so recent) proposals about non-commutativity of string zero 
modes \cite{leigh,Sakamoto:1989ig}.

 A key guide  in our analysis is the field theory description of 
gauge symmetry enhancement on toroidal compactifications. Gauge symmetry 
enhancement is a very stringy phenomenon associated to the fact that the string 
is an extended object and, therefore, it can wind around non-contractible 
cycles.
 At certain moduli points (i.e., fixed points of T-duality transformations) 
vector 
boson states, associated to definite  values of windings and compact  
momenta  become massless. These vectors,  combined 
with massless vectors inherited from the metric and antisymmetric tensor 
fields, 
  give rise to an enhanced gauge symmetry group $G_1$ (see for instance 
\cite{nsw,Giveon:1994fu}). 
 Further displacements on moduli space can lead  to  a  
different fixed point where, generically, other vectors associated to 
different  values 
of winding and momenta will become massless leading to a different enhanced 
gauge group $G_2$, etc. At generic points only a $ U(1)_L^{r+16}\times 
U(1)_R^r$
symmetry exists. Here,   
$r$ is the number of compactified dimensions  
associated 
to the KK zero modes of the metric and antisymmetric fields and  the 16 
comes from Cartan generators of the ten dimensional gauge group, 
in the heterotic string case.
The low energy effective  theory, {\it at a given moduli point}, where 
massive states are neglected, can be described by a usual gauge field theory  
Lagrangian coupled to  gravity  with no  explicit reference to any 
windings. By slightly  moving away from this fixed moduli point,  
gauge 
symmetry gets broken. The symmetry breaking can be understood as a 
conventional higgsing mechanism and also,  as found from a DFT approach 
\cite{aamr,aamp}, as associated to a dependence on moduli fields of 
the ``will-be structure 
constants'' fluxes.

The main aim of the present work is to write down a lower dimensional  
field theory able to provide a description of the enhancement phenomena 
occurring  on toroidally compactified heterotic string. This action  depends on 
moduli fields expectation values such that the 
different low energy effective field theories, associated to  heterotic 
enhancement situations, can be reached by varying such values.
 Our construction is restricted to fields corresponding to low string 
oscillator 
and includes   the fields that are involved 
in the 
enhancement phenomena. Clearly a full, consistent   description of the string 
theory  would require the introduction  of an infinite number of fields of all 
possible spins. We comment on a possible step by step completion of our 
construction, going beyond low energy, at the end of the article.

Very schematically, the idea is to incorporate a vector boson field $A_{\mu}(x, 
{\mathbb Y})$ and a scalar 
$M_{\bar{I}}(x,\mathbb{Y})$ into the action,  in addition to  the
fields inherited from the  usual ten dimensional metric, the dilaton and the 
Kalb-Ramond $B_2$. 
All fields must depend on both $d$ space-time $x^{\mu}$ 
coordinates as well as on internal compact toroidal   $\check{\mathbb 
Y}\equiv (y^I, y^m,\tilde y_m )$ coordinates. Namely, besides the $y^I$ 
coordinates
associated to the heterotic string degrees of freedom, $2r$ double coordinates 
$(y^m,\tilde y_m)$, conjugate to momenta and windings modes $(p_m,\tilde 
p^m)$,  for each of 
the $r$ compact dimension are considered in the spirit of DFT. 
A generalized mode expansion (GKK)  in periodic internal coordinates 
would produce $d$ dimensional fields $A_{{\nu}}^{({\mathbb 
L})}(x)$ (and $M_{{\bar{I}}}^{({\mathbb 
L})}(x)$)  with ${{\mathbb L}}$ labeling modes,  depending on windings 
and KK momenta. As mentioned before, for certain moduli  values some of these 
modes 
become massless 
and, when 
 combined with KK zero modes coming from metric and $B$ field (as well as 
heterotic Cartan fields) they enhance the gauge 
symmetry. The other modes, not participating in the enhancement process, remain 
very massive (with masses of the order of string mass ${\ap}^{-1}$  ) and  
do not contribute to the low energy effective theory.

The resulting action, in terms of the ``uplifted'' $A_{\mu}(x, {\mathbb Y})$ and 
$M_{\bar{I}}(x,\mathbb{Y})$
fields, appears to require a non-commutativity on fields introduced through 
a  non-commutative $\star$-product in 
the compact space \cite{leigh}. 
At the neighborhood of each specific  moduli fixed point and when only  
the slightly  massive  modes that become massless at this point are 
kept, 
the usual,  commutative, effective gauge theory   action is 
recovered  after integrating over the internal coordinates. The gauge 
symmetry gets enhanced   exactly at the fixed point.

Therefore, the 
action provides an effective interpolation among theories at 
different points. It is worth mentioning that enhancement can be described 
in DFT constructions as an enlargement 
of the compactification tangent space 
\cite{agimnr,aamr,Cagnacci:2017ulc,aamp, Fraiman} 
at a fixed point. Here, however,  the compact manifold is an $r$ dimensional 
double 
torus and  we find  that this enlargement is 
effectively provided by 
Fourier modes associated to fields that ``will-be  massless at such 
point''. 
Interestingly enough, the mentioned  non-commutativity can be traced back 
to cocycle factors in string vertices. These factors were first 
mentioned in \cite{Hull:2009mi} but did not manifest in previous DFT 
constructions due to the considered  level matching conditions and to the fact 
that calculations were  performed up to third order terms in the fields.

We organize the article as follows: In Section \ref{sec:The 
effective action}, we introduce 
the proposed action in $D=d+2r$ dimensions.
In Section \ref{sec:The action for GKK modes} we perform the mode expansion and 
analyze the different contributions. Section \ref{sec:Gauge symmetry 
breaking-enhancement along moduli space} deals with the 
physical content of the action, like  vector and scalar masses, 
Goldstone bosons, enhancement-breaking of gauge symmetries, etc. An  
illustrative  torus compactification ($r=2$) example  is briefly discussed.
A summary and a discussion of the limitations and possible extensions of the 
present work are presented in Section \ref{sec:Summary and Outlook}.
Notation and technical aspects are reserved to the appendices where  a 
more 
detailed description of the $\star$-product, is extended to incorporate the 
heterotic 
string gauge modes.

\section{The effective action}
\label{sec:The effective action}
In this section we present a  moduli dependent field 
theory effective action that captures  essential  features of 
symmetry enhancement in toroidal compactification of 
heterotic string. 
The  basic ingredients and notation conventions are introduced here. The  
reader is referred to the appendices for details. 

Let us denote by $\Phi\equiv(g,b,A) $ a moduli point  encoding the background  
metric $g$, the $b$ field and 
Wilson 
line values. At a given fixed point $\Phi_0$ on 
moduli space the heterotic gauge group is 
of the form $G_L\times U(1)^r_R$.
The rank of  $G_L$ is  $r_L=r+16=26-d$ originating in 
the $16$ Cartan 
generators of the ten dimensional heterotic gauge  group plus the $r=10-d$ 
vector bosons coming from  Left combinations  of the KK  reductions of 
the metric and the antisymmetric tensor. Therefore, the dimension of the gauge 
group is  $\dim{G}_L=n_c+r_L$ where $n_c$ denotes the number of charged 
generators. These generators correspond to string vertex operators 
containing KK momenta and 
windings associated, generically, with massive fields that  become massless 
at the fixed point.
These fields will play a central role in our construction.
Let us stress that  $n_c$ depends on the moduli point and that, at 
generic points, there is no enhancement at all ($n_c=0$) and the  generic gauge 
group  
is $U(1)^{r_L}_L\times U(1)^r_R$.
The low energy effective action for the bosonic sector of  
heterotic string,  at a fixed point $\Phi_0$ with $G_L\times U(1)^r_R$ gauge 
group 
 and up to third order in the fields, reads  
\bea \label{effectiveSD}\nn
S_{eff}(\Phi_0)&=& \int d^dx\sqrt{g}\Bigg[e^{-2\varphi}
\left(
R+4\partial^\mu\varphi\partial_\mu\varphi-\frac1{12}H_{\mu\nu\rho}H^{\mu\nu\rho
}\right) \nn\\ &-&\frac 14 \left({F}_{A\mu\nu}{F}^A_{\mu\nu} +
{ F}_{\bar J}^{\mu\nu} {  F}^{\bar  J}_{\mu\nu} -2 
g_d\sqrt{\ap} 
M_{A\bar I} 
F^A_{\mu\nu} F^{\bar I\mu\nu }\right) \nn\\
 &-&
\frac 14 D_\mu M_{A\bar I}D_\nu M^{A\bar I} g^{\mu\nu}
 +
 {\cal
 O}(M^4)\Bigg]
 \label{hetactionfp}
\eea
where $\bar I$ Right indices correspond to the Abelian Right group  
$U(1)_{R}^r$ and   $A$ indices label Left $G_L$ (generically non Abelian)  
group.
We have  
 \bea \label{FHDFT}
F^B & = & d A^B +  \frac{g_d}{2}f_{CD}{}^B A^C \wedge A^D, \qquad
F^{\bar I}  =  d A^{\bar I}\\
D_\mu M_{A\bar I}& = &\partial_\mu M_{A\bar I}+
 g_d f^K{}_{SA} A_{L\mu}^S
     M^{{}}_{K\bar I}\quad 
\eea
where scalar fields $ M_{A\bar I}$ live in  the $(\bf {dim G_L})_{\bar q =0}$ 
adjoint 
representation of $G_L$ and carry zero vector charge  ${\bf {\bar q }}=(\bar 
q_1,\dots,\bar q_r)=0$ with respect to $U(1)_{R}^r$ Abelian Right group. 
  $H$ is the $B$  field strength (with Chern-Simons interactions) defined as
  \begin{equation}
  H=dB + F^{B}\wedge A_{B},
  \end{equation}
   $\varphi$ is the dilaton and $R$ the 
scalar curvature. 

As mentioned above,   the terms in this expression corresponding 
to   fields  originating in reductions of 
the 
10D fields will be always present,  whereas terms associated with charged 
fields will change when moving on moduli space. 
In our construction it proves convenient  to separate these contributions and 
rewrite the above action \eqref{hetactionfp} as 
\bea \label{hetactionsplit}\nn
S_{eff}(\Phi_0)&=&\int d^dx\sqrt{g}e^{-2\varphi}\Bigg[
\left(
R+4\partial^\mu\varphi\partial_\mu\varphi-\frac1{12}H_{\mu\nu\rho}H^{\mu\nu\rho
}\right) \\\nn &-&\frac 14 \left(\delta_{\hat I\hat J}{F}^{\hat 
I\mu\nu}{F}^{\hat J}_{\mu\nu} + 
\delta_{\bar I\bar J}
{ F}^{\bar I\mu\nu} { F}^{\bar I}_{\mu\nu} -2 g_d\sqrt{\ap} 
M_{\hat I\bar I} 
F^{\hat I}_{\mu\nu}  F^{\bar I\mu\nu }+ D_\mu M_{\hat I\bar I}D_\nu 
M^{\hat I\bar I} g^{\mu\nu}\right) \nn\\
 &-&\frac 14 \left({F}_{\alpha\mu\nu}{F}^{\alpha}_{\mu\nu} +D_\mu 
M_{\alpha\bar I}D_\nu M^{\alpha\bar I} g^{\mu\nu}-
 2 g_d\sqrt{\ap} 
M_{\alpha\bar I} 
F^{\alpha}_{\mu\nu}  F^{\bar I\mu\nu }\right)  +
 {\cal
 O}(M^4)\Bigg].\nn \\
 \eea

Several indices are introduced: 
\begin{itemize}
 \item  $\bar 
I=1,\dots r$ 
are Right indices that label the Abelian 
group 
$U(1)^{\bar I}$  associated to Right vector bosons $  A_{\mu}^{\bar I}$ 
  \item The Left index $A$ has been conveniently splitted 
 as  $A=(\hat I,\alpha )$ where :
 
 $\alpha =1,\dots, n_c$  label the 
Left gauge group  
 charged 
generators 
with vector bosons $ A_{\mu}^{\alpha }$. They correspond to roots of the 
algebra 
in a Cartan-Weyl basis. 

$\hat I\equiv (I, m)$  are  Left  indices 
splitted in terms of $m=1\dots r$ compact Left indices and $ 
I=1\dots 16$ heterotic indices.     $ 
A_{\mu}^{\hat I}$ correspond to the Left Cartan vector boson fields.
\end{itemize} 
 The  field strengths  introduced in \eqref{FHDFT} 
are now splitted as 
\begin{eqnarray}
F^{\hat I}_{\mu\nu}  =&2\partial_{[\mu}A^{\hat I}_{\nu]} 
+i g\sum_{\alpha}f_{ 
\alpha \beta}{}^{\hat I} A^{\alpha}_{[\mu} A^{\beta}_{\nu]}\qquad F^{\bar 
J}_{\mu\nu}  = 2\partial_{[\mu}A^{\bar J}_{\nu]} 
\label{cartanusual}
\end{eqnarray}
 for Cartan fields whereas
\begin{eqnarray}
F^{\alpha}_{\mu\nu} & = &2\partial_{[\mu}A^{\alpha}_{\nu]} +igf^{\alpha}{}_{ 
\beta \gamma} A^{\beta}_{[\mu} A^{\gamma}_{\nu]} +igf^{\alpha}{}_{ \beta \hat I 
} 
A^{\beta}_{[\mu} A^{\hat I}_{\nu]}
\label{chargedusual}
\end{eqnarray}
are the field strengths for charged vectors\footnote{Here we use the convention 
$2 A_{[\mu} B_{\nu]}=A_{\mu} B_{\nu}-B_{\nu}A_{\mu}$.}.
Similarly, for scalar fields we have
\begin{eqnarray}
 D_\mu M_{\hat I\bar J}=\partial_\mu M_{\hat I\bar J}+
i g_d f_{\beta \hat I}{}^{\alpha} A_{\mu}^{\beta}
     M^{{}}_{\alpha \bar J}\, 
  \label{nonchargedscalarusual}  
\end{eqnarray}
\begin{eqnarray} D_\mu M_{\alpha \bar J}=\partial_\mu M_{\alpha \bar J}+
i g_d f_{\alpha \beta }{}^{\hat I}  M^{{}}_{\hat I \bar J} A_{\mu}^{\beta}
    + i g_d f_{ \alpha\beta}{}^{\lambda} M^{{}}_{\lambda \bar J}
A_{\mu}^{\beta}
    + i g_d f_{ {\alpha} \hat I}{}^{\beta }M^{{}}_{\beta \bar J}
A_{\mu}^{\hat I}
        \label{chargedscalarusual}
  \end{eqnarray}  
where a sum over repeated root indices is implicit.  We are using a Cartan-Weyl 
basis such that $f_{\alpha\beta}{}^{\gamma}=f_{\alpha\beta(-\gamma)}=1$ (with 
$\gamma=\alpha + \beta$) and $f_{\beta \alpha}{}^{\hat I}=f_{-\alpha 
\alpha}{}^{\hat I}=f_{\alpha-\alpha \hat{I}}=\alpha^{\hat{I}}$ (no sum on 
$\alpha$ here) etc. Also, charged indices are contracted with the corresponding 
Cartan-Killing form whereas Cartan indices contract with a delta function.

Finally, it proves useful to perform a further rewriting 
of the above action by collecting Left and Right   
``Cartan'' indices into a unique generalized  ${\cal I}= ({\hat I},\bar I)$ 
index spanning the vector representation of $O(r_l,r)$ duality group.
The $\cal I$ indices are contracted with an  $O(r_L,r)$ invariant metric 
that we will generically  express in the L-R basis (also called $C-$basis) as
 \be \label{etac}
\eta^{{\cal I} {\cal J}}_C=\begin{pmatrix} 1_{16+r} & 0 \\ 0 & -1_r 
\end{pmatrix}.
\ee
In order to have a covariantly looking form in this basis  we introduce the   
generalized vector $A^{\cal 
I}_{\mu}=
(A^{\hat I}_{\mu},A^{\bar I}_{\mu}) $ that incorporates the Left and Right 
Cartan 
fields respectively and define the scalars  $M_{\alpha \mathcal{J}}=(0, 
M_{\alpha \bar 
J}) $ where Left components are projected out.  We discuss this projection 
in \eqref{projectors} below. 
\
Also,  inspired by DFT constructions 
\cite{aamp} 
we introduce a generalized 
$O(r_l,r)$ 
 metric ${\cal H}_{}^{\cal I \cal J}$ and we expand on fluctuations 
around a flat background as 
\begin{equation}
 {\cal H}_{}^{\cal I \cal J}= \delta^{\cal I \cal J}+ {\cal H}^{(1)\cal I \cal 
J}+ \frac12{\cal 
H}^{(2)\cal I \cal J}+\dots
\label{scalarfluct}
\end{equation}
where matrix elements vanish  unless
\begin{eqnarray}
 {\cal H}^{(1)}_{\hat I\bar J}&=&- M_{\hat I\bar J},\qquad{\cal 
H}^{(1)}_{\bar J 
\hat I}=-M^T_{\hat I\bar J}\nn\\
{\cal 
H}^{(2)}_{\hat I\hat  J}&=& (M M^T)_{\hat I\hat  J},\qquad{\cal 
H}^{(2)}_{\bar I \bar J}= (M^T M)_{\bar I \bar J}\, .
\label{scalarorders}
\end{eqnarray}

In terms of this metric  and keeping terms up to 
second order in 
fluctuations (assuming vector fields are first order) the 
action\footnote{Notice that there is no scalar potential. Scalar 
interactions appear at fourth order in the fields when only massless states 
are considered.} can  be 
re-expressed  as 
 \bea\label{effectiveactionH}
 S_{eff}(\Phi_0) &=&
 \int d^dx\sqrt{g}e^{-2\varphi}\Bigg[
\left(
R+4\partial^\mu\varphi\partial_\mu\varphi-\frac1{12}H_{\mu\nu\rho}H^{\mu\nu\rho
}\right) \\\nn 
&-&\frac 14  
\mathcal{H}_{\cal I \cal J}{F}^{\cal 
I\mu\nu}{F}^{\cal J}_{\mu\nu} + \frac 
18 D_\mu 
\mathcal{H}_{\cal I\cal J}D^\mu  
\mathcal{H}^{\cal I\cal J} 
\\ \nn
 &-&\frac 14 {F}_{\alpha \mu\nu}{F}^{\alpha \mu\nu}  
 +\frac 14 D_\mu M_{\alpha \cal I}D^\mu M^{\alpha \cal I}-\frac12 
g_d\sqrt{\ap} 
M_{\alpha \cal I} 
F^{\alpha}_{\mu\nu} \bar F^{\cal I\mu\nu }\Bigg].
\eea

The different contributions to the action above are read out from 3-point 
amplitudes of massless  heterotic string vertices. Vertex operators for 
vector bosons and scalars  considered here are collected in Table 1 and 
explained below.

\begin{table}[h]\label{tablavertices}
\hspace{ 2 cm}
\begin{tabular}{|c|c|c|c| }
 \hline
& & &\\[-0.4cm]
Field Modes & $\mathbb{L}^2 $ & $N$ & Vertex Operators\\
\hline
  & & &\\
$g_{\mu\nu}^{(\mathbb{L})},b_{\mu\nu}^{(\mathbb{L})},\phi^{(\mathbb{L})}$
  & 0 & 1 & 
$\partial_{z}X^{\mu}\tilde{\psi}^{\nu}(z)e^{i\mathbb{L}.\mathbb{Y}(z)}e^{iK.X(z)
}$\\ \hline
& &  & \\
$A^{\bar{I} (\mathbb{L})}_{\mu}$  &  0  & 1 &  
$\partial_{z}X^{\mu}\tilde{\psi}_{\bar{I}}(z)e^{i\mathbb{L}.\mathbb{Y}(z)}e^{
iK.X(z)}
$\\ \hline
& &  & 
$\partial_{z}Y_{I}\tilde{\psi}^{\mu}(z)e^{i\mathbb{L}.\mathbb{Y}(z)}e^{iK.X(z)}
$\\[-0.3cm]
$A^{\hat{I}(\mathbb{L})}_{\mu}$  & 0   & 1 & \\[-0.2cm]
& &  
&$\partial_{z}Y_{m}\tilde{\psi}^{\mu}(z)e^{i\mathbb{L}.\mathbb{Y}(z)}e^{iK.X(z)}
$\\
 \hline
  & & &\\[-0.4cm]
  & & 
&$\partial_{z}Y^{I}\tilde{\psi}^{\bar{I}}(z)e^{i\mathbb{L}.\mathbb{Y}(z)}e^{
iK.X(z)}
$\\[-0.2cm]
  $M_{\hat{I}\bar{I}}^{(\mathbb{L})}$ & 0  & 1 & \\[-0.2cm]
  & &  
&$\partial_{z}Y^{m}\tilde{\psi}^{\bar{I}}(z)e^{i\mathbb{L}.\mathbb{Y}(z)}e^{
iK.X(z)}
$\\
 & &  &\\[-0.4cm]
 \hline
 &   &   &\\
$A_{\mu}^{\mathbb{(L)}}$ & 2    & 0  & 
$\tilde{\psi}^{\mu}(z)e^{i\mathbb{L}.\mathbb{Y}(z)}e^{iK.X(z)}$\\[0.2cm]
\hline
& &  &\\[-0.4cm]
$M^{(\mathbb{L})}_{\bar{I}}$ & 2 & 0  & 
$\tilde{\psi}^{\bar{I}}(z)e^{i\mathbb{L}.\mathbb{Y}(z)}e^{iK.X(z)}$\\[0.2cm]
 &    & & \\
\hline
\end{tabular} \hspace{2cm} 
\caption{\footnotesize Field Modes, LMC, oscillator number and corresponding 
string 
vertex operator. In all cases considered here $\bar N=\bar N_F+\bar 
N_B-\frac12=0 $.}
\end{table}
String vertex operators generically contain an internal factor (see Appendix 
\ref{sec:Heterotic string 
basics} for notation)
\begin{eqnarray}
 e^{i\mathbb{L }^{(\check {\mathbb P})}(\Phi).{\mathbb Y}(z)}= 
e^{il_L^{(\check {\mathbb P})}.{y}_L(z) +i l_R^{(\check {\mathbb P})}.y_R(z)}
  \end{eqnarray}
where $\mathbb{L }^{(\check {\mathbb P})}(\Phi)=
(l_L^{(\check {\mathbb P})}( \Phi),l_R^{(\check {\mathbb P})}(\Phi))$ is 
the 
generalized momentum\footnote{We 
will generally write $\mathbb{L }\equiv \mathbb{L }^{(\check {\mathbb 
P})}(\Phi)$ and  omit the explicit writing of the 
dependence on  $\check {\mathbb P}$ and $\Phi$ to lighten  the 
notation.} (see \eqref{generalizedmomentum}) that depends  on 
windings  $\tilde p^{m}$,  KK momenta $p_{m}$ and $\Lambda_{16}$ weights 
$P^I$ that we organize into  the generalized Kaluza-Klein (GKK) 
momenta 
\begin{equation}                                     
              {\check 
{\mathbb P}}\equiv (P^I ,p_{m},\tilde p^{m} ) 
\label{generalizedKK}
         \end{equation}                                     
   and on moduli field values $\Phi$ .  Generalized momenta are constrained to 
satisfy the Level Matching 
Condition (LMC) 
\begin{equation}
 \frac12{\mathbb L}^2=\frac12 l_L^2- \frac12l_R^2=\tilde 
p.p+\frac12P^2=(1-N+\bar N).
\label{LMCwidingstext}
\end{equation}
In all the cases considered here $\bar N=0$ and $N=0,1$. Vertex operators with 
$N=1$ correspond to KK reductions of the metric, B-field,  dilaton field and 
heterotic vector fields in 
10 dimensions.

For instance, the Cartan vectors  $A_{\mu}^{\hat I}$   do originate in 
string vertex operators coming from KK reductions of the metric and 
antisymmetric field of the form 
\begin{eqnarray}
  V({\hat I},\mathbb{L})\propto A^{\hat{I}(\mathbb{L})}_{\mu}(K)\, 
\partial_z 
Y_{{\hat I}} \tilde\psi^{\mu} e^{i\mathbb{L }^{(\check {\mathbb 
P})}(\Phi).{\mathbb Y}(z)}e^{iK.X(z)}         
 \end{eqnarray}
where $K^{\mu}$ is the space time momentum. Due to the presence of 
oscillators $ \partial_z 
Y^{{\hat I}}$, $N=1$ and therefore LMC \eqref{LMCwidingstext} 
reads  \begin{equation}\label{lmccartan}
\mathbb{L}^2=0.
       \end{equation}
 This requirement is trivially satisfied by massless states that 
correspond 
to 
$\mathbb{L}\equiv l_L=l_R=0$ (with null windings and KK momenta), as it is 
indeed the case for  Cartan vectors $A_{\mu}^{\hat I}\equiv 
A^{(0)}_{\mu\hat{I}} $.   
 
On the other hand, the left handed charged vector bosons arise  from 
vertices 
\begin{eqnarray}\label{vertexbosoncharged}
 V({{\mathbb L}})\propto  A_{\mu}^{({\mathbb L})}(K) \,
\tilde\psi^{\mu}(z) e^{i\mathbb{L }^{(\check {\mathbb P})}(\Phi).{\mathbb 
Y}(z)}
  e^{iK.X(z)}
\end{eqnarray} with LMC
\begin{equation}\label{lmccharged}
\frac12{\mathbb{L}^2}=1
       \end{equation}
since $N=0$. The other cases in the Table 1 are  understood in a similar 
way. Let us 
stress that ghost factors  as well as cocycle factors must be included.

At a fixed point $\Phi_0$ and for specific values of windings and momenta 
(i.e., for specific values of  $\check {\mathbb 
P}$)  
\begin{eqnarray}
 l_R^{(\check{\mathbb{P}})}( \Phi_0)=0 \qquad {\rm  } \qquad 
l_L^{(\check{\mathbb{P}})}(\Phi_0)=\alpha^{(\check{\mathbb{P}})} \qquad {\rm 
with }  \qquad \frac12\alpha^{(\check{\mathbb{P}})2}=1
 \label{masslescondfp}
\end{eqnarray}
the states become massless (see \ref{LRstringmasses})  and   
$l_L^{(\check{\mathbb{P}})}(\Phi_0)$ become the 
roots $\alpha^{(\check{\mathbb{P}})}$ of the 
enhanced gauge group algebra  charged generators\footnote{As string theory 
operators,  
 $e^{il_L^{(\check{\mathbb{P}})}.Y(z)}\rightarrow 
J_{\alpha^{(\check{\mathbb{P}})
}}$ are the 
charged generators of the algebra.}. Generically, at a different fixed point,  
other set of  
$\check{\mathbb{P}}'$s will ensure \eqref{masslescondfp}, leading to a 
different enhanced gauge group. 
We will denote this set of $n_c$ GKK modes, satisfying 
\eqref{lmccharged}, 
by 
\begin{equation}
 \check G(\Phi_0)_{n_c}=\{\check{\mathbb{P}}\equiv (P^I, 
p_{m},\tilde p^{m}):l_R^{(\check{\mathbb{P}})}( \Phi_0)=0 \,({\rm thus}\quad 
l_L^{(\check{\mathbb{P}})}(\Phi_0)=\alpha^{(\check{\mathbb{P}})}, m^2=0)\}.
\label{enhancementsector}
\end{equation}
Namely,  $\check G(\Phi_0)_{n_c}$ encodes the $n_c$ ``will-be massless charged 
fields at fixed point $\Phi_0$''.
At $ \Phi_0$, and for ${\check {\mathbb P}}\in\check G(\Phi_0)_{n_c}$  
the $A_{\mu}^{(\check {\mathbb P})}(K)$ modes give rise to charged vector field 
 $A_{\mu}^{\alpha}(x)$ in the action above (similarly with charged scalars).

As stated in the Introduction the main aim of our work is to provide a 
unified field theory description such that at 
given fixed points the different effective gauge theories are reproduced.
 Following the suggestions 
in \cite{aamr} we propose to consider a sort of 
generalized Kaluza-Klein   expansion on generalized momenta $\mathbb  L$ 
of the different 
fields coming into play in the enhancement process. The GKK modes in this 
expansion are identified with a corresponding polarization  of a vertex 
operator. For instance, in order to describe charged vector bosons we introduce 
the expansion
\bea
A_{\mu}(x, {\mathbb Y})
=\sum_{\check {\mathbb 
P}}'
A_{{\mu}}^{( {\mathbb 
L})}(x) e^{i \mathbb{L}_{\cal I} \mathbb{Y}^{\cal I}}\,
=\sum_{{\mathbb L}}
A_{\mu}^{({\mathbb L})}(x) e^{i l_L.y_L+i l_R.y_R}\,\delta 
(\frac12\mathbb  L^{ 2},1)
\label{vectorexp}
\eea
where  $A_{{\mu}}^{({\mathbb L})}(x)$ correspond to polarization modes in 
\eqref{vertexbosoncharged}. The prime in the sum indicates that LMC  
\eqref{lmccharged} must be imposed (with an abuse of notation we indicate the 
sum 
on  
mode index  $\check {\mathbb P}$ by  $\mathbb L$).  Recall that 
generically the sum contains an infinite number of  terms even though  the LMC 
is a 
severe constraint
.

Generically, if the mass of the GKK 
components $A_{{\mu}}^{({\mathbb L})}(x)$ were  given by the 
string mass formula \eqref{LRstringmasses}, as we will show to be the case, 
these modes 
would  be massive. However,  when moving continuously along the moduli 
space, for specific values
$\check {\mathbb P}\in \check G_{n_c}(\Phi_0)$,  $n_c$ vector fields
$A_{\mu}^{({\mathbb L})}(x)\equiv A_{\mu}^{\alpha^{(\check {\mathbb P})}}(x)$
would  become massless and would lead to  the enhanced $G_L$ gauge 
group\footnote{A  reality condition   
$A_{{\mu}}^{({\mathbb L})*}=A_{{\mu}}^{(-{\mathbb L})}$ must be imposed.}.
In a similar way we introduce the GKK expansion for scalar fields by 
associating the fields $M_{\alpha \bar I}(x)$, coming from string vertex 
operators modes $M_{\bar I}^{({\mathbb L})}(K)$ (see Table 1 above) to modes in 
a GKK expansion 
\begin{eqnarray}
M_{\bar I}(x, {\mathbb Y})=\sum_{{\mathbb L}} M_{\bar I}^{({\mathbb L})}(x) 
e^{i \mathbb{L}_{\cal 
I} \mathbb{Y}^{\cal 
I}}\, \delta 
(\frac12\mathbb  L^{ 2},1).
\label{scalarexp}
\end{eqnarray}
Before addressing the other mode expansions let us note that the  Right 
fields $M_{\bar I}$ can be embedded into constrained fields 
$M_{\cal I}= (M_{\hat I},M_{\bar I})$ with $O(r_L,r)$ indices. Namely, they are 
defined as 
\begin{eqnarray}\label{projectors}
  P_{\cal I}{}^{\cal J} M_{\cal J} =0   \qquad    P&=&\frac12(\eta+{\cal 
H})\\\nn
  \bar P_{\cal I}{}^{\cal J} M_{\cal J}= M_{\cal I} \qquad 
\bar P &=&\frac12(\eta-{\cal H})
\end{eqnarray}
where $\cal H$ is the generalized metric satisfying ${\cal 
H}\eta{\cal H}=\eta$ and  $P, \bar P$ are projectors \cite{dftheterotic} that 
eliminate $r_L$  degrees of 
freedom. From the  first equation we obtain, by plugging in the 
generalized metric expansion \eqref{scalarorders},

\begin{equation}
  P M=\frac12(\eta+{\cal H})M=
\begin{pmatrix} (1 + M M^T)_{\hat I\hat  J}+\dots & - M_{\hat 
I\bar J} +\dots\\ -(M^T)_{\bar I\hat 
J}+\dots  & -(M^TM)_{\bar I \bar J} +\dots
\end{pmatrix}\begin{pmatrix}M_{\hat J}\\M_{\bar 
J}\end{pmatrix}=0
\end{equation}
where $\dots $ indicate higher order terms in fluctuations. Therefore $M_{\hat 
I}= - M_{\hat I\bar J}M_{\bar 
J}+\dots$.
We see that  $M_{\hat I}$ degrees of freedom are not independent and contribute 
at order two or higher in fluctuations. As we show below, these will give rise 
to terms of order four in the action. For this reason  we 
can set $M_{\cal J}=(0, M_{\bar J})$.

Expansions of fields  originating in the $D=10$ metric, $B$ field 
and the Cartan generators 
of the heterotic gauge group, namely, $G_{\mu\nu}(x, {\mathbb Y}), 
B_{\mu\nu}(x, 
{\mathbb Y})$, $ A_{\mu}^{\cal I}(x, {\mathbb Y}),M^{\hat I \bar J}(x, 
{\mathbb 
Y})$ must also be considered. Now, since the corresponding modes (first four 
rows in Table 1)  must satisfy LMC \eqref{lmccartan} we restrict the sum to 
modes obeying this constraint. For instance
\begin{eqnarray}
  A_{\mu}^{\cal I}(x, {\mathbb Y})=\sum_{{\mathbb L}}  A_{\mu}^{\cal 
I ({\mathbb L})}(x) 
e^{i \mathbb{L}_{\cal 
I} \mathbb{Y}^{\cal 
I}}\, \delta 
(\mathbb  L^{ 2},0).
\label{cartanbosonexp}
\end{eqnarray}

Recall that these modes correspond to $N=1$ and, therefore, only the zero mode 
 $A_{\mu}^{{\cal I} (0)}(x)=A_{\mu}^{{\cal I}}(x)= (A^{\hat I}_{\mu}(x),A^{\bar 
I}_{\mu}(x)) $ would correspond to  massless fields. These are the vector 
fields of the 
$U(1)^{r_L}_L\times U(1)^r_R$ gauge group for a generic point in moduli space.
The same considerations are valid for the other $N=1$ fields. Thus, for 
example,  $G_{\mu\nu}(x)^{(0)}=G_{\mu\nu}(x)$ is the $d$ dimensional metric 
field whereas non-zero modes would describe massive gravitons, etc.
In most of the considerations below only these zero modes   will 
be 
needed. 

Notice that a generic, moduli dependent,  field $\phi(x, {\mathbb Y})\equiv 
A_{\mu}(x, {\mathbb Y}), G_{\mu\nu}(x, {\mathbb 
Y}),\dots $ could be 
interpreted as an uplifting of $d$ dimensional fields to  $d+r+r_l$ 
dimensions with  $r+r_l$ periodic\footnote{Recall that 
Heterotic coordinates can be thought of as  coordinates on a 16 
dimensional torus with a chiral projection.}. A Lagrangian ${\cal L}(x, 
{\mathbb Y})$ in terms of these 
fields, when integrated over the $d+r+r_L$ dimensions $\int d^dx \,
d\mathbb{Y}\, {\cal L}(x, {\mathbb Y})=\int d^dx\, dy_L\, dy_R\,{\cal L}(x, 
{\mathbb Y})$ will lead to an action in $d$ space-time dimensions after 
periodic coordinates are integrated out, where the physical fields will be the  
$\phi(x)^{(\mathbb{L})}$ GKK modes.  Our expectation is  that such 
action includes the 
effective low energy heterotic effective action \eqref{effectiveactionH} for 
different fixed moduli point $\Phi_0$.

A crucial point is how to 
generate  a non-Abelian  structure out of these fields in order to give rise to 
 enhancements at fixed points. We will see that the job is accomplished by a new 
so called ``star 
product'' \cite{leigh}, which we denote by $\star $,  
accounting for  non-commutativity. 

In the following we  present the action and subsequently we  discuss 
its particular 
features.
Let us assume that we are able to write down  a full field theory action $ 
S_{het}(\Phi)$  by computing all possible  heterotic string theory amplitudes.
 This action  should include 
an infinite number of fields, let us call them $\Phi_{\mu_1\mu_2\dots;\bar 
N,N}(x,\mathbb Y)$, of all possible spins and oscillator numbers $\bar 
N,N $  that must 
be mode expanded with the corresponding level matching condition $\frac12 
{\mathbb L}^2=1-(N-\bar N)$.  
Among all these contributions we  isolate  the action piece, 
that we call $S_{enh}(\Phi)$,   
containing   up to third order terms (and some fourth order as we 
discuss below) and   involving  fields coming form $10D$ KK reductions 
$G_{\mu\nu},B_{\mu\nu}, A^{\bar J}_{\mu}, \varphi,M_{\hat I,\bar 
J},A^{\hat  
J}_{\mu} $ and the extra fields $A_{\mu}, M_{\bar I} $. 
These  fields  are associated 
to oscillator numbers  $N=0,1$, respectively, and $ \bar 
N=0$. Their  corresponding modes are 
collected in Table 1 as well as their associated string vertex operators. 

Therefore, we split  the full action into 
\begin{equation}
 S_{full \,het}(\Phi)=S_{enh}(\Phi)+S'(\Phi)
 \label{fullaction}
\end{equation}
where the term  $S'$ encodes all other (infinite)  contributions that we 
are not explicitly considering here. These include  higher spin 
fields, fields associated to oscillator numbers $N> 1$, higher order terms in 
fluctuations, etc. 
 $S_{enh}(\Phi)$ is the action we are going to deal with,  given by 
\begin{equation}
\begin{aligned}
S_{enh}(\Phi)=&\int d^dxd\mathbb{Y}\sqrt{g} e^{-2\varphi}\Bigg[
(
R+4\partial^\mu\varphi\partial_\mu\varphi-\frac1{12}H_{\mu\nu\rho}H^{\mu\nu\rho
}) \\
& -\frac{1}{4}
\mathcal{H}_{\mathcal{I}\mathcal{J}}\star F^{\mathcal{I}}_{\mu\nu}\star 
F^{\mathcal{J} \mu\nu}  + 
\frac{1}{8}\mathcal{D}_{\mu}\mathcal{H}^{\mathcal{I}\mathcal{J}}\star\mathcal{D}
^{\mu}\mathcal{H}_{\mathcal{I}\mathcal{J}}\\
&-\frac{1}{4} F_{\mu\nu}\star F^{\mu\nu}+\frac{1}{4} 
\mathcal{D}_{\mu}M^{\mathcal{I}}\star\mathcal{D}^{\mu}M_{\mathcal{I}}-\frac{1}{2
} M_{\mathcal{I}} \star F^{\mu\nu}\star F^{\mathcal{I}}_{\mu\nu}  \\
& - 
\frac{1}{4} 
\partial_{\mathcal{J}}M^{\mathcal{I}}\star\partial_{\mathcal{K}}M_{\mathcal{I}}
\big(H^{\mathcal{J}\mathcal{K}} -\eta^{\mathcal{J}\mathcal{K}})+ i \frac{1}{2} 
\partial_{\mathcal{I}}M^{\mathcal{J}}\star 
M_{\mathcal{J}}\star M^{\mathcal{I}}
\Bigg].
\end{aligned} 
 \label{hetactionuplift}
\end{equation}
As we have already emphasized the different terms in the action are 
expressed in terms of 
the fields 
introduced above. 
These fields depend on the compact 
coordinate $\mathbb{Y}$ and 
can therefore be mode expanded. 
Integration over $\mathbb{Y}$ will produce  
an effective action in $d$ space-time dimensions.
In the next section we perform the mode expansions and integrate over 
internal coordinates in order to obtain a $d$-dimensional space time 
action. Before presenting these computations let us first discuss the 
general structure and the kind of  information we expect  this action 
to contain. 

Fields originating in $D=10$ KK 
reductions, i.e. $G_{\mu\nu},B_{\mu\nu},\varphi, M_{\hat I,\bar J},
A^{\cal J}\equiv (A^{\hat  
J}_{\mu}, A^{\bar J}_{\mu})$, 
require a mode expansion with the constraint ${\mathbb L}^2=0$ whereas fields 
$A_{\mu},M_{\cal I}$, associated to enhancements,  require 
$\frac12{\mathbb L}^2=1$.
It appears somewhat unnatural  to indicate what kind of constrained mode 
expansion must be performed in each case. However, these LMC constraints 
might  be implemented in the Lagrangian through, for instance, Lagrange 
multipliers.  Thus, if we indicate by $\phi_N(x, {\mathbb Y})$ a field such 
that its mode expansion must be restricted to $\frac12{\mathbb L}^2=1-N$, in a 
DFT language we  would require
\begin{equation}
-\frac12  \partial_{\cal I}\partial^{\cal I} \phi_N(x, {\mathbb Y})= 
-\frac12(\partial_{L}^2-\partial_{R}^2) \phi_N(x, {\mathbb Y})=1-N
\label{lmcmultiplier}.
\end{equation}
 In the cases considered here  $N=0,1$ label  the number of Left 
indices.
Clearly, 
the so called \textit{strong constraint} of DFT (see for 
instance \cite{Hull:2009mi, reviewamn}) cannot be satisfied if enhancement 
phenomena are included. 

The term $\frac{1}{2}(\eta^{\mathcal{J}\mathcal{I}} - 
H^{\mathcal{J}\mathcal{I}})$ acts as a covariant $O(r_l,r)$ projector.

If just the zero modes are kept we notice that  the first two rows 
 in \eqref{effectiveactionH} are formally reproduced with 
$g_{\mu\nu}=G_{\mu\nu}^{(0)},M_{\hat I\bar J}=M_{\hat I\bar 
J}^{(0)} $, etc.  However, a non trivial action of 
the $\star$-product arises whenever  non zero  modes come into play as it 
happens, for instance, in products  of fields associated to enhancements (and 
thus 
requiring expansions with 
$\delta(\frac12{\mathbb L}^2,1)$ constraint). We  provide a more 
detailed   discussion of this situation   in the next section.
Also, the different terms in the action are now defined as 
\begin{equation}
\begin{aligned}
& F_{\mu\nu} = 2 \partial_{[\mu}A_{\nu]} + ig A_{\mu}\star A_{\nu} + 
2gA^{\mathcal{I}}_{[\mu}\star\partial_{\mathcal{I}}A_{\nu]}\\
& F_{\mu\nu}^{\mathcal{I}} = 2 \partial_{[\mu}A^{\mathcal{I}}_{\nu]} + 
g\partial^{\mathcal{I}} A_{\mu} \star A_{\nu}+4ig A_{[\mu} \star 
A_{\nu]}^{\mathcal{I}}
\label{Fuplift}
\end{aligned}
\end{equation}

\begin{equation}
\begin{aligned}
& \mathcal{D}_{\mu}M^{\mathcal{I}} = \partial_{\mu}M^{\mathcal{I}} + ig 
A_{\mu}\star M^{\mathcal{I}} + g 
A^{\mathcal{J}}_{\mu}\star\partial_{\mathcal{J}}M^{\mathcal{I}} - g 
\partial_{\mathcal{J}}A_{\mu}\star\big( H^{\mathcal{J}\mathcal{I}} - 
\eta^{\mathcal{J}\mathcal{I}} \big)\\
& \mathcal{D}_{\mu}\mathcal{H}^{\mathcal{I}\mathcal{J}} = 
\partial_{\mu}\mathcal{H}^{\mathcal{I}\mathcal{J}} + g 
\partial^{\mathcal{I}}A_{\mu}\star M^{\mathcal{J}}+2i g 
A_{\mu}\star\mathcal{H}^{\mathcal{I}\mathcal{J}}+ 
i g 
A_{\mu}^{\mathcal{K}}\,\partial_{\mathcal{K}}\mathcal{H}^{\mathcal{I}\mathcal{J}
} 
\end{aligned}
\label{covderivativesuplift}
\end{equation}
(where $g=\frac{1}{\sqrt{\alpha'}}$) by generalizing 
\eqref{chargedusual}. 

Finally, the three form $H$ is defined as
\begin{equation}
H=dB + F^{\mathcal{I}}\star\wedge A_{\mathcal{I}} + F\star\wedge A.
\end{equation}

Whenever a product of two fields appears a $\star$-product must be used. 
For instance, the generalized metric must be expressed in terms of 
fluctuations as in \eqref{scalarfluct} but with  a $\star$ replacing 
the ordinary product.

All the  fields that we are  considering contain modes that are 
massless at some  specific values of  moduli $\Phi$. This is always the case 
for the zero modes  $G_{\mu\nu}^{(0)}(x),B_{\mu\nu}^{(0)}(x),\varphi^{(0)}(x), 
M_{\hat I,\bar J}^{(0)}(x)$ 
that are massless everywhere in moduli space, whereas $n_c$ modes  
$A_{\mu}^{(\check {\mathbb P})}(x),  M_{\bar I}^{(\check {\mathbb 
P})}(x)$ become massless at a point $\Phi_0$ for momenta in $
 \check G(\Phi_0)_{n_c}$ (see \eqref{enhancementsector}). These are the modes 
that participate in the enhancement phenomena. When approaching a point 
$\Phi_0$ in  moduli space the light spectrum will contain the zero 
mode massless fields plus the $n_c$ slightly massive  modes in $
 \check G(\Phi_0)_{n_c}$, all other fields  having 
masses of the order of the string mass. When moving to some other 
fixed point $\Phi_1$ other set of modes (intersections can 
occur) in 
$\check G(\Phi_1)_{n_c}$ will become light\footnote{There will always be 
modes that remain very massive, as for instance $G_{\mu\nu}^{(\mathbb 
L)}(x) $ with $\mathbb 
L\ne 0$.}.    Therefore, the action 
\eqref{hetactionuplift} can be splitted as 
\begin{equation}
 S_{enh}(\Phi)=S_{{\rm light\,at}\,\Phi_0 }(\Phi)+S_{{\rm 
heavy\,at}\,\Phi_0}(\Phi)=S_{{\rm light\,at}\,\Phi_1 }(\Phi)+S_{{\rm 
heavy\,at}\,\Phi_1}(\Phi)=\dots
 \label{enhancesplit}
\end{equation}
The first (second) splitting is convenient when $\Phi$  is close to $\Phi_0$ 
($\Phi_1$). In this case, at low energies,  the second term in the action (and 
also $S'$ above), containing heavy states (of order ${\ap}^{-1}$) and light 
states in interaction with them,  does  not 
contribute. We  will be left with the effective $S_{eff}(\Phi\approx 
\Phi_0)$ low energy action 
 \begin{equation}\label{leeffa}
  S_{eff}(\Phi\approx \Phi_0)=S_{enh}(\Phi\approx \Phi_0)=S_{{\rm 
light\,at}\,\Phi_0 }(\Phi\approx \Phi_0)
 \end{equation}
and similarly for $\Phi\approx \Phi_1$ etc. 
At $\Phi= \Phi_0$ all fields in   $S_{eff}(\Phi_0)$ become massless and the 
effective action should reproduce \eqref{hetactionsplit} with 
gauge group $G_L\times U(1)^r_R$. The $\star$-product plays a crucial role in 
reproducing the  non-Abelian group structure.  When slightly moving away 
from $\Phi_0$, the gauge symmetry should break, generically,  to $ 
U(1)_L^{r+16}\times 
U(1)_R^r$ 
and 
$S_{eff}(\Phi\approx \Phi_0)$ should contain massless and massive 
physical states correctly transforming under the Abelian groups.

Besides these features  addressed  in the next section when mode expansions 
are performed, we stress that $S_{enh}(\Phi)$ appears to encode 
some relevant information about very massive states as discussed in an explicit 
example in \ref{subsec:$SU(3)$ example}.

\section{The action for GKK modes}
\label{sec:The action for GKK modes}
In this section we perform the expansion of  the fields in the above action in 
terms of GKK modes,  compute the $\star $-products for these modes and 
finally integrate over the internal coordinates $\mathbb{Y}$ in order to obtain 
the moduli dependent $d$ dimensional effective action. In particular, we 
will show that  after integrating out the massive modes, the  massless 
GKK modes at a self-dual 
point,  Eq.\eqref{hetactionuplift} give rise to the gauge enhanced 
action \eqref{effectiveactionH}. 

The particular $\star$-product we consider here is a generalization of the one 
 proposed in \cite{leigh} to the case of the heterotic string. It is 
described in Appendix \ref{sec:The star product}. For
two 
mode expanded fields it reads 
\begin{eqnarray}\label{star2fields}
(\phi_{N_1}\star\psi_{N_2})(x,\mathbb{Y})&=&\sum_{\mathbb{L}_1,\mathbb{L}_2}'
e^{i\pi 
{l}_1\cdot \tilde l_2} 
\phi^{(\mathbb{L}_1)}_{N_1}(x)\psi^{(\mathbb{L}_2)}_{N_2}(x)e^{i(\mathbb{L}_1 + 
\mathbb{L}_2).\mathbb{Y}} \\\nn\end{eqnarray}
where  a phase ${l}_1\cdot \tilde l _{2}={p}_{1 m}\tilde p^{2m}+{p}_{1I} 
\tilde p_{2}^I$ 
dependent on the KK momenta  ${l}_1$  of the first field and the windings
$\tilde l_{2}$ of second mode is generated  (see \eqref{mwheterotic}).
The first term corresponds to a sum over the internal  compactification lattice 
indices. The sum over heterotic directions is  constrained by a chiral 
projection that eliminates Right heterotic momenta. It  can be 
expressed as 
\begin{equation}\label{so32phases}
 p _{1I}\tilde{p}_2^I= \frac12 P_1EP_2
\end{equation}
 in terms of  $\Spin(32)$,\footnote{We will mainly refer to $\Spin(32)$ but 
results are valid 
for  $E_8\times E_8$ as well.} weights and 
$E_{IJ}=G_{IJ}+B_{IJ}$ (see Appendix \ref{sec:The star product}). The prime in 
the sum 
indicates that the constraint $\frac12 
\mathbb{L}_i^2=1-N_i$ $(i=1,2)$ must be imposed for the field with subindex 
$N_i$.
By using that
\begin{equation}\label{phase}
l_1\cdot\tilde{l}_2  
 + l_2 \cdot \tilde{l}_1
=\mathbb{L}_1\mathbb{L}_2= \frac12 
(\mathbb{L}_1+\mathbb{L}_2)^2-
\frac12\mathbb{L}_1^2-\frac12\mathbb{
L} _2^2
\end{equation}
by recalling that the exponents are just integer multiples of $\pi$ and by 
using LMC  we can rewrite the above product as
\begin{eqnarray}
(\phi_{N_1}\star\psi_{N_2})(x,\mathbb{Y}) &=&\sum_{\mathbb{L}_1,\mathbb{L}_2}'
e^{i\pi (\frac12\mathbb{L}^2-N_1-N_2)}
e^{i\pi l_2 \cdot \tilde{l}_1}
\cdot 
\phi^{(\mathbb{L}_1)}_{N_1}(x)\psi^{(\mathbb{L}_2)}_{N_2}(x)e^{i(\mathbb{L}_1 + 
\mathbb{L}_2).\mathbb{Y}} 
\label{starheterotic}
\end{eqnarray}
where $\mathbb{L}=\mathbb{L}_1+\mathbb{L}_2$. 
By comparing to \eqref{star2fields} we see that the $\star$-product is non 
commutative unless the phase 
$e^{i\pi (\frac12\mathbb{L}^2-N_1-N_2)}=1$.
For instance,  if $\mathbb{L}=0$, as 
we would find if we integrated on $\mathbb{Y}$, we find 
\begin{equation}
 \int d\mathbb{Y} \,\phi_{N_1}\star\psi_{N_2}=e^{i\pi 
(N_1+N_2)} \int d\mathbb{Y} \,\psi_{N_2}\,\star\phi_{N_1}.
\end{equation}
For the cases we are considering here we notice that fields with similar LMC 
commute whereas for $N_1=0, N_2=1$ (or viceversa) they anticommute.
 
Let us proceed to consider the product of three  fields. 
In this case, by  using associativity (see \eqref{assocstarproduct}), we  have
\begin{eqnarray}
(\phi_{N_1}\star\psi_{N_2}\star 
\lambda_{N_3})(x,\mathbb{Y})&=&\sum_{\mathbb{L}_1,\mathbb{L}_2,\mathbb{L}_3 
}' \tilde{f}_{\mathbb{L}
_1 
\mathbb {L}_2
\mathbb{L}_3} 
\phi^{(\mathbb{L}_1)}_{N_1}(x)\psi^{(\mathbb{L}_2)}_{N_2}(x)\lambda^{(\mathbb{L}
_3) }_{ N_3} (x)e^ {
i(\mathbb{L}_1 + \mathbb{L}_2+\mathbb{L}_3).\mathbb{Y}}
\label{eqstarproduct3Fourier}
\end{eqnarray}
where
\begin{equation}\label{3fieldsphase}
 \tilde{f}_{\mathbb{L}
_1 
\mathbb {L}_2
\mathbb{L}_3}=e^{i{\pi } 
{l}_1\cdot \tilde l_2}e^{i{\pi } 
({l}_1+{l}_2)\cdot \tilde l_3}\equiv\pm 1.
\end{equation}
If  we integrate over internal 
coordinates, due to  momentum conservation $\mathbb{L}= \mathbb{L}_1 + 
\mathbb{L}_2+\mathbb{L}_3=0$, the second phase 
becomes $e^{-i\pi\frac12 \mathbb{L}_3^2}$ and therefore 
\begin{equation}\label{phaseL0}
  \tilde{f}_{\mathbb{L}_1 \mathbb {L}_2 
\mathbb{L}_3}=e^{i\pi{l}_1\cdot \tilde l_2}e^{-i\pi\frac12 
\mathbb{L}_3^2}
\end{equation}
In this case, 
\begin{eqnarray}\nn
  \tilde{f}_{\mathbb{L}_1 \mathbb {L}_2 \mathbb{L}_3} &=&
  e^{i\pi {l}_1 \cdot \tilde l_2} e^{-i\pi\frac12 
\mathbb{L}_3^2}= 
  e^{i\pi ({l}_1\cdot \tilde l_2+{l}_2\cdot \tilde l_1)}
  e^{i\pi{l}_2\cdot  \tilde l_1}e^{-i\pi\frac12 
\mathbb{L}_3^2}\\
&=&
e^{i{\pi }(\frac12 
\mathbb{L}_3^2-\frac12 
\mathbb{L}_2^2-\frac12 
\mathbb{L}_1^2)} e^{i\pi {l}_2\cdot \tilde l_1}e^{-i\pi\frac12 
\mathbb{L}_3^2}=-e^{i\pi (N_1+N_2+N_3)}\tilde{f}_{\mathbb{L}_2
\mathbb {L}_1 \mathbb {L}_3}
\label{chargessc}
\end{eqnarray}
 where we have used  \eqref{phase} above with
$\mathbb{L}_1+\mathbb{L}_2=-\mathbb{L}_3$.
A similar phase is obtained if $2\leftrightarrow 3$.

We conclude that  the product of three fields with $N_i=0$, as it is the case 
for charged fields participating in the enhancements, the phases   $ 
\tilde{f}_{\mathbb{L}_1 
\mathbb {L}_2
\mathbb{L}_3}$  are completely antisymmetric under index 
permutation. This result  is valid both for  massless 
and massive states.
On the contrary, for modes originating in 10D fields, $N_i = 1$, and the phase 
becomes irrelevant so the $\star$-product reduces to just the ordinary product.

In the following subsections we analyze the different 
contributions in the action \eqref{hetactionuplift} in terms of their mode 
expansions.

\subsection{The vectors kinetic term }
 Let us  analyze first $\int 
d^dxd\mathbb{Y}~F_{\mu\nu}\star 
F^{\mu\nu}$. In order to do it let us consider  the Fourier\footnote{We 
normalize the integration variables  so as 
to have a unit volume 
factor. Also, we use that
$
\int d^{2n}\mathbb{Y} e^{i 
(\mathbb{P}_{M}+\mathbb{Q}_{M}) \mathbb{Y}^{ 
M}}=\delta^{2n}(\mathbb{P}_{M}+\mathbb{Q}_{M})\, ,
$} 
component (see 
\eqref{Fuplift}) $F^{(\mathbb{L})}_{\mu\nu}$ as follows 
\begin{equation}
F^{(\mathbb{L})}_{\mu\nu} = \int\, d\mathbb{Y} F_{\mu\nu} 
e^{-i\mathbb{L}\cdot \mathbb{Y}} 
,
\end{equation}
where:
\begin{eqnarray}
 F_{\mu\nu}^{(\mathbb{L})}&=&2\partial_{[\mu}A^{(\mathbb{L})}_{\nu]} +
 ig\sum_{\mathbb{L}_2}'\tilde f_{\mathbb{L} \,
\mathbb {L}_2
\mathbb{L}_3}A_{\mu }^{
(\mathbb{L}_2)}A^{(\mathbb{L}_3)}_{\nu}\Bigg.
+2i g \sum_{\mathbb{L}_2}' f_{\mathbb{L}_3 
\mathbb {-L}_3
{\cal I}}A^
{{\cal I}(\mathbb{L}_2)}_{[\mu}A_{\nu]}^{(\mathbb{L}_3)}.
\label{modesfmunu}
\end{eqnarray}
Here, $\mathbb{L}_3 = \mathbb{L}-\mathbb{L}_2$, and  we have 
defined  
$\tilde{f}_{\mathbb{L} 
\mathbb {-L}
\mathcal{I}}$ as
\begin{equation}
\tilde f_{\mathbb{L} 
-\mathbb {L}
\hat I} =  l_{\hat I}(\Phi),\qquad \tilde f_{\mathbb{L} 
-\mathbb {L}
\bar I}= l_{R,\bar 
I}(\Phi).
\end{equation}\label{structureconstant}
%
 Recall that  $F_{\mu\nu}^{(\mathbb{L})}$ depends on moduli point.

Let us look at the contributions, at low energy,  at self-dual 
points. On the one hand, from the last sum we keep the zero mode contribution 
$A^
{{\cal I}(0)}_{\mu}\equiv A^{\mathcal{I}}_{\mu}(x)$ giving rise to Cartan 
vector fields. On the other hand  the light modes on the first sum correspond
to $\check{\mathbb P}\in \check G_{n_c}(\Phi_0)$. Thus, when sliding to  
$\Phi=\Phi_0$ 
 (\ref{modesfmunu}) will reduce to (\ref{chargedusual}) as long as we identify 
\begin{equation}\label{eqidentAnegroot}
A_{\mu}^{(\mathbb{L}_i)} \leftrightarrow A_{\mu}^{\alpha^{(\mathbb{L}_i)}},~~~ 
-A_{\mu}^{(-\mathbb{L}_i)} \leftrightarrow A_{\mu}^{\alpha^{(-\mathbb{L}_{i})}},
\end{equation}
where $\mathbb{L}_i$ is in a one to one correspondence with the positive roots 
$\alpha_i\equiv \alpha^{(\mathbb{L}_i)}$, $i=1,...,\frac{n_c}{2}$, (and 
$-\mathbb{L}_i$ with $-\alpha_i\equiv \alpha^{(-\mathbb{L}_i)}$) of the enhanced 
group (the underlying  reason for this identification is  
$f_{\mathbb{L}\mathbb{L}_2\mathbb{L}_3}$ is invariant under 
$\mathbb{L}_i\rightarrow -\mathbb{L}_i$).  
Thus, for each root  $\alpha^{({\check{\mathbb 
P})}}=l_L^{(\check{\mathbb{P}})}(\Phi_0) $ (see \eqref{masslescondfp})
\begin{eqnarray}\label{fschargedgenerators}
F_{\mu\nu}^{\alpha^{({\check{\mathbb P})}}}&=&
2\partial_{[\mu}A^{\alpha^{({\check{ \mathbb P})}}}_{\nu]} +
 ig\sum_{{{\check{ \mathbb P_2 
}}}}'f_{
 {\alpha^{({\check{ \mathbb P}_2)}}}
{\alpha^{({\check{ \mathbb P}_3)}}}{\alpha^{({-\check{ \mathbb P})}}}}A_{\mu 
}^{\alpha^{({\check{ \mathbb P}_2)}}}A^{{\alpha^{({\check{ \mathbb 
P}_3)}}}}_{\nu}\Bigg.
+ 2igf_{{\hat I} {\alpha^{({\check{\mathbb P})}}} {\alpha^{({-\check{\mathbb 
P})}}}
}A_{[\mu}^{{\hat I}}A^
{{\alpha^{({\check{\mathbb P})}}}}_{\nu]}\\\nn
F_{\mu\nu}^{\alpha^{({-\check{\mathbb P})}}}&=&
2\partial_{[\mu}A^{\alpha^{({-\check{ \mathbb P})}}}_{\nu]} +
 ig\sum_{{{\check{ \mathbb P}_2}}}'f_{
 {\alpha^{({-\check{ \mathbb P}_2 )}}}
{\alpha^{(-{\check{ \mathbb P}_3)}}}{\alpha^{({\check{ \mathbb P})}}}}A_{\mu 
}^{\alpha^{({-\check{ \mathbb P}_2)}}}A^{{\alpha^{(-{\check{ \mathbb 
P}_3)}}}}_{\nu}\Bigg.
+ 2igf_{{\hat I} {\alpha^{({-\check{\mathbb P})}}} {\alpha^{({\check{\mathbb 
P})}}}
}A_{[\mu}^{{\hat I}}A^
{{\alpha^{({-\check{\mathbb P})}}}}_{\nu]}
\end{eqnarray}
where we have identified  $f_{
 {\alpha^{(-{\check{ \mathbb P}_1)}}}
{\alpha^{({\check{ \mathbb P}_2)}}}{\alpha^{({\check{ \mathbb P}_3)}}}}
 =f_{-\alpha_1 \alpha_2 
\alpha_3}=-f_{\alpha_1 -\alpha_2 -\alpha_3}$ as the algebra structure 
constants for 
charged generators. Therefore, \eqref{fschargedgenerators} becomes the field 
strength for the charged fields   of the corresponding gauge 
theory. Then, up to third order in fluctuations
we can write
\begin{align}\label{chargedk}\nonumber
\int dxd\mathbb{Y}~F_{\mu\nu}\star F^{\mu\nu}  = &-\sum_{\mathbb{L}}'\int 
dx 
F_{\mu\nu}^{(\mathbb{L})}F^{(\mathbb{-L}) \mu\nu}\\
& = \int dx F_{\alpha\mu\nu}F^{\alpha \mu\nu}.
\end{align}
We have thus matched the first term of the third row of action 
\eqref{effectiveactionH}. The second term of the same row is 
reproduced by $\mathcal{H}_{\mathcal{I}\mathcal{J}}\star 
F^{\mathcal{I}}_{\mu\nu}\star 
F^{\mathcal{J} \mu\nu}$ in \eqref{hetactionuplift} since, when focusing 
only on massless GKK modes, 
\begin{equation}\label{cartanfs}
F^{\mathcal{I}}_{\mu\nu} = 2\p_{[\mu}A^{\mathcal{I}}_{\nu]} + 
2ig\sum_{\check{\mathbb{P}}}f_{\alpha^{(\check{\mathbb{P}})}\alpha^{(-\check{
\mathbb{P}})}}{}^{\mathcal{I}} A^{\alpha^{(\check{\mathbb{P}})}}_{[\mu} 
A^{\alpha^{(-\check{\mathbb{P}})}}_{\nu]}
\end{equation}
(remember 
$f_{\alpha^{(\check{\mathbb{P}})}\alpha^{(-\check{\mathbb{P}})}}{}^{\bar I}=0$ 
for massless GKK modes).
\subsubsection{$D=10$ heterotic string action}
 It is interesting to consider the $D=10$ theory.   In this case 
 $\mathbb{L }\equiv l_L$ is always Left handed where  $ 
l^I=P^I$ are just the components of the $\Spin(32)$ roots 
$P=(\underline{ \pm 1,\pm 1,0,\dots})$ (underlining denoting permutation). 
Therefore,  \eqref{chargedk} becomes 
the  $\Spin(32)$ gauge kinetic term for charged fields. On the other 
hand  Eq.\eqref{cartanfs} with ${\cal I}=1,\dots 16$ provides the field 
strength for the Cartan components and, therefore,  when combined with the 
other  terms in second and first rows 
in
\eqref{hetactionuplift}  the low energy $D=10$ heterotic 
effective action is 
recovered.
Recall that none of the other terms are present since there is 
no compactification at all.\

\subsection{Scalars kinetic term}
\label{sec:Charged scalars kinetci term}
Following similar steps as above we can write
\begin{eqnarray}\nonumber
\int dxd\mathbb{Y}~\mathcal{D}_{\mu}M_{\cal J}\star \mathcal{D}^{\mu}M^{\cal J}
 =-\int dx~\sum_{\mathbb{L}}'\mathcal{D}_{\mu}M^{(\mathbb{L})}_{\bar J}
\mathcal{D}^{\mu}M^{(\mathbb{-L})\bar J}
 \end{eqnarray}
 where
 \begin{eqnarray}
\label{covderivativemode}
  \mathcal{D}_{\mu}M^{(\mathbb{L})}_{\bar 
J}&=&\partial_{\mu}M^{(\mathbb{L})}_{\bar J}+
  ig\sum_{\mathbb{L}_2}'\tilde{f}_{\mathbb{L} \mathbb {L}_2
\mathbb{L}_3}M^{(\mathbb{L}_2)}_{\bar J}A_{\mu}^{ (\mathbb{L}_3)}+ 
 ig\tilde{f}_{\mathbb{L} \mathbb {-L}
I} M^{(\mathbb{L})}_{\bar J} A_{\mu}^{I(0)}\\\nn
&+&ig\sum_{\mathbb{L}_2}'
\tilde{f}_{\mathbb{L} \mathbb {L}_2\mathbb{L}_3}
M_{\hat I \bar J}^{(\mathbb{L}_2)}
\mathbb{L}_3^{\hat I} A_{\mu}^{(\mathbb{L}_3)}+ igl_{R\bar{I}} 
A_{\mu}^{\bar{I}(0)}M^{(\mathbb{L})}_{\bar 
J}-2igl_{R\bar{J}} 
A_{\mu}^{(\mathbb L)}+\dots
 \end{eqnarray}
 is the Fourier transform of first equation in  \eqref{covderivativesuplift} 
with $\mathbb {L}_3= 
\mathbb {L}- \mathbb {L}_2$. We have used that 
 $(\bar{H}_{\cal I \cal J}-\eta_{ \cal I  
\cal J})\mathbb{L}^{ \cal J}=(0,2\,l_{R\bar{I}})$. The  dots amount for 
terms 
containing massive modes $A_{\mu}^{I(\mathbb{L}_3)}$.
The usual covariant derivative for charged vectors of the gauge group $G_L$ 
(first term of the 
second row of \eqref{effectiveactionH}) is reproduced at enhancement point 
$\Phi_0$,  for $\check{\mathbb P}\in \check G_{n_c}(\Phi_0)$ with the 
identifications   
\begin{equation}
M_{\bar{J}}^{(\mathbb{L}_i)} \leftrightarrow 
M_{\bar{J}}^{\alpha^{(\mathbb{L}_i)}},~~~ -M_{\bar{J}}^{(-\mathbb{L}_i)} 
\leftrightarrow M_{\bar{J}}^{\alpha^{(-\mathbb{L}_{i})}}.
\end{equation}

Covariant derivative of scalar  modes $M_{\hat I\bar{J}}^{(\mathbb{L}_i)}$ arise 
from  the  term 
$\frac{1}{8}\mathcal{D}_{\mu}\mathcal{H}^{\mathcal{I}\mathcal{J}}\star\mathcal{D
}^{\mu}\mathcal{H}_{\mathcal{I}\mathcal{J}}$ in \eqref{effectiveactionH} and 
read 
\begin{eqnarray}\nn
\label{covderivativeKKscalarmode}
  \mathcal{D}_{\mu}M^{(\mathbb{L})}_{\hat I \bar 
J}&=&\partial_{\mu}M^{(\mathbb{L})}_{\hat I\bar J}+
ig\sum_{\mathbb 
{L}_2}'\tilde{f}_{\mathbb{L} 
\mathbb{L}_2 \mathbb{L}_3}A_{\mu}^{ 
(\mathbb{L}_2)}\mathbb{L}_{2{\hat I}} M^{(\mathbb{L}_3)}_{ 
\bar J}
+2ig\sum_{l}'\tilde{f}_{\mathbb{L} 
\mathbb{L}_2 \mathbb{L}_3} A_{\mu}^{ (\mathbb{L}_2)}
 M_{\hat I  \bar J}^{(\mathbb{L}_3)}\\
&+&  ig\mathbb{L}_{\hat K} A_{\mu}^{{\hat K}(0)}M^{(\mathbb{L})}_{\hat I\bar J} 
       +igl_{R\bar{I}} 
A_{\mu}^{\bar{I}(0)}M^{(\mathbb{L})}_{\hat I\bar J}+...
 \end{eqnarray}
 The massless scalars are provided by the zero modes $ M_{\hat I  \bar 
J}^{(0)}$. In this case, the  last two 
terms drop out and  \eqref{nonchargedscalarusual} is reproduced.
\subsection{Scalar potential and other couplings}
\label{sec:Scalar potential}
The scalar potential is such that it vanishes (up to third order in 
fluctuations)  for massless states, it is 
$O(r_L,r)$ invariant and it reproduces the scalar potential away from the 
enhancing point for scalars that would be massless at such point, as computed 
in \cite{aamp}. It appears 
that the most general form is 
\begin{eqnarray}
\frac{1}{4} 
\partial_{\mathcal{J}}M^{\mathcal{I}}\star\partial_{\mathcal{K}}M_{\mathcal{I}}
\big(\mathcal{H}^{\mathcal{J}\mathcal{K}} -\eta^{\mathcal{J}\mathcal{K}} \big) 
+ i 
\frac{1}{2} \partial_{\mathcal{I}}M^{\mathcal{J}}\star M_{\mathcal{J}}\star 
M^{\mathcal{I}}
+ \mathcal{O}(M^{4}).
 \label{scalarpotential}
\end{eqnarray}

It is worth noticing that the first term 
above,  when opened up in terms of fluctuations, contains a 
fourth-order term (at the enhancement point) of the following schematic form: 
$\sum_{\alpha}f_{I\alpha 
-\alpha}f_{J-\alpha\alpha}M_{I\bar{I}}M_{J\bar{I}}M_{\alpha\bar{K}}M_{ 
-\alpha\bar{K}}$ and is  part of the fourth order scalar 
potential \cite{Fraiman}. 
To 
complete the full fourth-order terms in the potential more terms are needed.  
For instance we would need an extra term of the form 
$\partial_{\mathcal{I}}\phi_{\bar{I}}\star\partial^{\mathcal{I}}\phi_{\bar{I}} 
\star\phi_{\bar{J}}\star\phi_{\bar{J}}$ among others. We leave the analysis of 
these extra terms for future work.

Finally our action \eqref{hetactionuplift} contains a last term, 
 $M_{\mathcal{I}} \star F^{\mu\nu}\star F^{\mathcal{I}}_{\mu\nu}$, which 
gives rise to the last term of \eqref{effectiveactionH} at the enhancement 
point 
(actually this term  is always present and away from the  self-dual point 
it gives rise to the 
adequate coupling between a massive scalar, a massive  vector and a massless 
$U(1)_{R}$ 
vector).
\section{ Breaking and enhancement of gauge symmetry along  moduli space}
\label{sec:Gauge symmetry breaking-enhancement along moduli space}
We have shown the explicit mode expansions for some of the terms appearing 
in the action. Computation of other terms proceed by following similar steps.

Several interesting results like vector and scalar masses, presence of 
would-be Goldstone bosons, etc.   can be straightforwardly  read out from these 
expansions. 
We discuss some of these issues below.

\subsection{Vector masses}
\label{sec:vector masses}
Vector boson masses can  be extracted from the fourth term in the covariant 
derivative-like
term \eqref{covderivativemode}
\begin{equation}
\begin{aligned}
\int dxd\mathbb{Y}~\frac{1}{4} D_{\mu}M_{\mathcal{J}}\star 
D^{\mu}M^{\mathcal{J}}&=\sum'_{\mathbb{L}} -\frac{1}{4} 
D_{\mu}M^{(\mathbb{L})}_{\bar J}D^{\mu}M^{(\mathbb{-L})\bar J}=\\
&=...+\sum'_{\mathbb{L}}\frac{1}{4}g^{2}2\big[\mathbb {L}\cdot\big(\bar{H}-\eta 
\big)\cdot\mathbb {L} \big] A_{\mu}^{(\mathbb{L})}
A_{\mu}^{(\mathbb{-L})}\\
&=\dots 
+\sum'_{\mathbb{L}} \frac{m^{2}_{A}}{2} 
A_{\mu}^{(\mathbb{L})}A_{\mu}^{(\mathbb{-L}) }
\end{aligned}
\end{equation}
where we have used the string theory result 
$m^2_A=2 l_R^2=\mathbb {L}\cdot\big(\bar{H}-\eta \big)\cdot\mathbb {L}$ for the 
masses of the charged vector fields.  We also observe  that there are no mass 
terms for Cartan vectors 
$A^{ \cal I}_ {\mu}$ so, generically, the gauge group is $U(1)^{r_L} \times 
U(1)^r$. 
At  $\Phi=\Phi_0$ and for  $\check{\mathbb P}\in \check G_{n_c}(\Phi_0)$ 
vector bosons become massless leading to gauge symmetry enhancement.

 Finally we  check the normalizations. Since the  kinetic term of 
the vectors reads:
\begin{equation}
-\int dxd\mathbb{Y}~\frac{1}{4} F_{\mu\nu}\star 
F^{\mu\nu}=\sum'_{\mathbb{L}}\frac{1}{4}\big(\p_{\mu}A^{(\mathbb{L})}_{\nu}
-\p_
{\nu}A^{(\mathbb{L})}_{\mu}\big)\big(\p_{\mu}A^{(-\mathbb{L})}_{\nu}-\p_{\nu}A^{
(-\mathbb{L})}_{\mu}\big) +...
\end{equation} 
when adding \eqref{covderivativemode} we find 
\begin{equation}
\sum'_{\mathbb{L}}\frac{1}{4}\big(\p_{\mu}A^{(\mathbb{L})}_{\nu}-\p_{\nu}A^{
(\mathbb{L})}_{\mu}\big)\big(\p_{\mu}A^{(-\mathbb{L})}_{\nu}-\p_{\nu}A^{
(-\mathbb{L})}_{\mu}\big) + \frac{m^{2}_{A}}{2} 
A_{\mu}^{(\mathbb{L})}A_{\mu}^{(\mathbb{-L}) }
\end{equation}
which is the Proca Lagrangian with signature $(-+++...)$ (with a global 
different normalization) of a vector with mass $m_{A}$.

\subsection{Goldstone bosons}
\label{sec:Goldstone bosons}
From the same scalars kinetic factors  above we find the terms
\begin{equation}
\begin{aligned}
+\sum'_{\mathbb{L}}\frac{1}{4} D_{\mu}M^{(\mathbb{L})}_{\bar J}
D^{\mu}M^{(\mathbb{-L})\bar J}&=...+\sum'_{\mathbb{L}}2\frac{1}{4}\mathbb 
{L}^{\bar 
J}\partial_{\mu} M^{(\mathbb{L})}_{\bar J}A_{\mu}^{(\mathbb{-L})}\\
&=
...+\sum'_{\mathbb{L}}\frac{1}{2} l_R^{\bar 
J}\partial_{\mu} M^{(\mathbb{-L})}_{\bar J} A_{\mu}^{(\mathbb{L})}.
\end{aligned}
\end{equation}
As also discussed in \cite{aamp} this coupling indicates that, for a given 
vector boson  $A_{\mu}^{(\mathbb{L})}$, there exists a  combination 
of $ \bar I=1,\dots,r=10-d$  of  would-be Goldstone bosons  $l_R^{\bar 
J}\partial_{\mu} M^{(\mathbb{-L})}_{\bar J}$ (this is exactly the Goldstone 
combination which can be read from the vertex operators in string theory 
\cite{agimnr}).
In fact, at enhancement point $\Phi _0$ and  for  $\check{\mathbb P}\in 
\check G_{n_c}(\Phi_0)$ these terms are not present since $l_R=0$. However,  
when sliding away from $\Phi _0$, $l_R\ne 0$, and these $n_c$ combinations 
provide the longitudinal components of the $n_c$ corresponding  $ A_{\mu}  
{(\mathbb{L})} $ massive vector bosons. Namely, 
\begin{equation}
  A_{\mu}^  
{(\mathbb{L})'}= A_{\mu}^  
{(\mathbb{L})}+l_R^{\bar 
J}\partial_{\mu} M^{(\mathbb{L})}_{\bar J}.
\end{equation}
\subsection{Scalar masses}
\label{sec:scalar masses}
The masses of the scalar fields can be read from the quadratic terms in the 
scalar potential \eqref{scalarpotential}  
\begin{eqnarray}\nn
 &&\frac14 \int dxd\mathbb{Y}(\mathcal{H}^{\cal I \cal J}-{\eta}^{\cal I \cal 
J}) \partial_{\cal 
I} M^{\cal K}\star \partial_{\cal 
J}M_{\cal K}=\sum_{\mathbb{L}}'\frac{1}{4}\big[\mathbb{L}\cdot\big(\bar H - 
\eta\big)\cdot\mathbb{L}\big] M^{(\mathbb{L})\cal K} M^{(-\mathbb{L})}_{\cal 
K}+\dots\\
&=&\sum_{\mathbb{L}}'\frac{1}{4}m^2_{M^{(\mathbb{L})}} 
M^{(\mathbb{L})\bar K} M^{(-\mathbb{L})}_{\bar K}+\dots
\end{eqnarray}
with 
\begin{equation}
m^2_{M^{(\mathbb{L})}}=2l_R^2=\mathbb{L}\cdot\big(\bar H - 
\eta\big)\cdot\mathbb{L},
\end{equation}
as expected from string theory. They coincide with the masses of the 
corresponding vector boson modes. 

 As in the vector case, it is still necessary to check for the relative  
coefficients, so we 
must compare the above terms with the scalar kinetic term
\begin{equation}
\int d\mathbb{Y}~\frac{1}{4} D_{\mu}M_{\cal J}\star D^{\mu}M^{\cal 
J}=-\sum'_{\mathbb{L}}\frac{1}{4}\p_{\mu}M^{(\mathbb{L})\cal 
J}\p^{\mu}M^{(-\mathbb{L})}_{\cal J}+\dots
\end{equation}
Altogether we have
\begin{equation}
-\sum'_{\mathbb{L}}\left[\frac{1}{4}\p_{\mu}M^{(\mathbb{L})\cal 
J}\p^{\mu}M^{(-\mathbb{L})}_{\cal J}+ \frac{1}{4}m^2_{M_{\bar 
J}^{(\mathbb{L})}} M^{(\mathbb{L})\cal J} M^{(-\mathbb{L})}_{\cal J}\right]
\end{equation} 
which is the Lagrangian of a massive scalar field (with a global normalization) 
 on the signature $(-+++...)$ with mass $m_{M_{\bar J}^{(\mathbb{L})}}$.
\subsection{Duality and gauge invariance}
Let us close this section  by commenting on T-duality and gauge invariance.
 
We notice that, even if the different 
terms in the action \eqref{effectiveactionH} are written in an $O(r_L,r)$ 
invariant way, by index contraction, the effect of 
the $\star$-product on T-duality is not transparent. On the one hand we expect 
that, whenever windings and momenta are included, the symmetry group becomes 
$O(r_L,r,{\mathbb 
Z})$. Consider  for instance 
\eqref{starheterotic}. Each term on the expansion 
contains a Fourier mode labeled by  $\check{\mathbb P}$,
encoding momenta and winding modes \eqref{generalizedKK}, as well as 
an exponential term $e^{i\mathbb L.\mathbb Y}$ where the exponent is 
explicitly $O(r_L,r,{\mathbb 
Z})$ covariant.
On the other hand, a $h \in O(r_L,r,{\mathbb 
Z})$ rotation leads to  
$\check{\mathbb 
P}\rightarrow \check{\mathbb P}'=h\check{\mathbb 
P}$ but also to  $e^{i\pi 
\tilde{l}_1l_2}\phi^{(\check{\mathbb 
P}_1)}(x)\psi^{(\check{\mathbb P}_2)}(x)\rightarrow 
e^{i\pi 
\tilde{l}'_1l'_2}\phi^{(\check{\mathbb 
P}_1')}(x)\psi^{(\check{\mathbb P}_2')}(x)$, where the heterotic part is 
expressed in terms of 
16 windings and momenta  as discussed in \eqref{mwheterotic}. 
However, if 
$\check{\mathbb 
P}$ satisfies LMC so does $\check{\mathbb 
P}'$ and, since we are summing over all modes satisfying LMC, the sum 
will be invariant.

Notice that, if we restricted  our analysis to  a set  of GKK momenta in 
$\check 
G(\Phi_0)_{n_c}$, the above transformations will take us out of this set. 
Namely, the $\check{\mathbb 
P}'$ will not become massless at  $\Phi_0$. However, $\check{\mathbb 
P}' \in \check 
G(h\Phi_0)_{n_c}$, and therefore their associated fields  will 
become massless at the transformed moduli 
point $h\Phi_0$ (note that the mass terms are invariant when transforming both 
the background and momenta and windings).
We will illustrate this fact in an example below.
Let us stress that if any of  the fields contained an   
$O(r_L,r)$ index, as for example $M_{\cal I}^{(\mathbb P)}$, it must appear 
contracted in an 
invariant way as it indeed happens in the action.

The action we are dealing with contains 
massive and massless states. At 
a $U(1)^{r_L} \times U(1)^r $ generic points, besides the $2r+16$ Abelian 
vectors and the gravity sector fields, all other vector and scalar  fields 
will be massive.  Recall that a field $\Phi(x)^{({\mathbb L})}$ carries charge 
$ {({\mathbb L})}_{\cal I}=  \tilde{f}_{\mathbb{L} \mathbb {-L}
\cal I} $ with respect to the Abelian factor  $A_{\mu}^{(\cal I)}$. 
Therefore,  under a $U^{\mathcal{I}}(1)$ gauge transformation 
\begin{eqnarray}
 \Phi(x)^{(\mathbb{L})}&\rightarrow & e^{ig\mathbb{L}_\mathcal{I} 
\alpha^{\mathcal{I}}(x)} \Phi(x)^{(\mathbb{L})}\\\nn
A^{\mathcal{I}}_{\mu}&\rightarrow &  A^{\mathcal{I}}_{\mu} 
-\partial_{\mu}\alpha^{\mathcal{I}}
\end{eqnarray}
and therefore, gauge invariance should 
be ensured by a  derivative of the form 
$D_{\mu}\Phi(x)^{({\mathbb 
L})}=\partial_{\mu}\Phi(x)^{({\mathbb L})} +i{({\mathbb L})}_{\cal I}
A_{\mu}^{(\cal I)}\Phi(x)^{({\mathbb L})}$. In fact, it can be checked that 
this is indeed the case for the covariant 
derivative of scalars in Eq. \eqref{covderivativemode} as well as for the 
derivative of the massive vectors as  given in Eq.\eqref{modesfmunu}. 
For instance, in the latter case we have that
\begin{align}\nonumber
F_{\mu\nu}^{(\mathbb{L})} & \rightarrow  
e^{ig\mathbb{L}_{\mathcal{I}}\alpha^{\mathcal{I}}(x)}2\partial_{[\mu}A^{(\mathbb
{L})}_{\nu]} + 
2ige^{ig\mathbb{L}_{\mathcal{I}}\alpha^{\mathcal{I}}(x)}\mathbb{L}_{\mathcal{I}}
A_{[\nu}\partial_{\mu]}\alpha^{\mathcal{I}} +  \\\nonumber
& ~~~~ + i g e^{ig\mathbb{L}_{\mathcal{I}}\alpha^{\mathcal{I}}(x)} 
\sum'_{\mathbb{L}_2}\tilde{f}_{\mathbb{L}\mathbb{L}_{2}\mathbb{L}_{3}}A_{\mu}^{
(\mathbb{L}_2)}A_{\nu}^{(\mathbb{L}_3)} +
2ige^{ig\mathbb{L}_{\mathcal{I}}\alpha^{\mathcal{I}}(x)}\tilde{f}_{\mathbb{L}
-\mathbb{L}\mathcal{I}}(A^{\mathcal{I}}_{[\mu} - 
\partial_{[\mu}\alpha^{\mathcal{I}})\mathbb{L}_\mathcal{I}A^{(\mathbb{L})}_{\nu]
}\\
& = 
e^{ig\mathbb{L}_{\mathcal{I}}\alpha^{\mathcal{I}}(x)}F_{\mu\nu}^{(\mathbb{L})}
\end{align}
where we have used momentum conservation $\mathbb{L}=\mathbb{L}_2+\mathbb{L}_3$.
Therefore  $F_{\mu\nu}^{\mathbb{(L)}}F^{\mu\nu \mathbb{(-L)}}$ terms in the 
action are $U(1)$ invariant.

On the other hand, at a given fixed point the Abelian gauge group is enhanced 
to some non-Abelian gauge group $G$, and all fields in the 
theory, massless and massive, should organize into $G$ irreducible 
representations. We have shown that, at a fixed point $\Phi_0$ or close to it, 
after very  massive states are integrated out (see discussion around 
\eqref{leeffa},  a well defined low energy effective gauge theory is obtained. 
The light modes that define this theory are the zero modes coming from 10D 
fields KK reductions plus modes in $ \check G(\Phi_0)_{n_c}$. 

However, if we were to consider  the other (very massive) modes, as the ones 
appearing in the mode summations we are dealing with,   we expect to  run 
into trouble. Indeed, generically, these massive modes will fill gauge 
multiplets that will contain modes associated to higher oscillator numbers. 
This appears as a limitation of our construction restricted to $N=0,1$ modes. 

Indeed, assume that ${\mathbb K}=(k_L,k_R)$ with $k_R=0$, $k_L^2=2$ 
encode the charged gauge vector boson modes $A_{\mu}^{( {\mathbb K})}$ of the 
group $G$ and let us call the currents associated to these vectors  $J^{( 
{\mathbb K})}$. From a string theory analysis, if we start with some massive 
field $\Phi(x)^{({\mathbb L})}$ with mass 
${m^{({\mathbb L})}}$ and  $\frac12{\mathbb L}^2=1$, its  OPE 
with the current should lead to another field $\Phi(x)^{({\mathbb S})}$ with 
${\mathbb S }= (s_L,s_R)= {\mathbb L}+ {\mathbb K}$ and the same mass, in order 
to 
be 
part of the same multiplet. Thus,  by using \eqref{leftrightmomenta} and 
\eqref{LMCwidings} we find that $\Phi(x)^{({\mathbb S})}$ mode should have 
$\bar N_s= \bar N_B+\bar 
N_F+\tilde 
E_0=0$ and a left oscillator mode $N_s$ such that
  \begin{eqnarray}\nn
   \frac{\ap}2 m^{2}_{L}&=& \frac12 l_L^2-1=
   \frac12 s_L^2+N_s-1 =\frac12 ({\mathbb L}+{\mathbb K})_L^2+ 
N_s-1=\frac12 l_L^2+   l_Lk_L+
N_s.
\label{multipletmasss}
  \end{eqnarray}
Namely 
\begin{eqnarray}
  l_Lk_L=-1 -
N_s \label{completeirrep}
\end{eqnarray}
  or, equivalently, $\frac12{\mathbb S }^2=1-N_s$. Therefore, even if we 
started  with a field corresponding to zero oscillator number we conclude that 
other values must be generically included. 
Presumably   full consistency would be 
attained if $\delta(\frac12{\mathbb S }^2,1-N)$ LMC is allowed in for all 
possible values of $N$. However, this 
would imply introducing higher spin fields, as expected form string theory.

Interestingly enough, it appears that  consistency (at tree level) can be 
achieved up to first mass level, with $ N=0,1$ as we are indeed considering 
here. We discuss this issue in the example below.

Finally recall that several consistency conditions  are 
expected to be satisfied by physical states. For instance, physical massive 
vectors must satisfy $\partial^{\mu} A^{B}_{\mu} = 0\,$, etc. 
In string theory such conditions arise from conformal invariance. Namely, 
physical fields must satisfy the adequate OPE with the stress energy tensor.
It was shown in Ref.\cite{amn}, in the case of the bosonic string and for some 
specific fields,   that these conditions can be understood from generalized 
diffeomorphism invariance. 
However, as mentioned above, when level matching conditions as  
$\frac12\mathbb L^2=1$ (or $\frac12\mathbb L^2=0$ for non zero modes) are  
considered  our  
analysis points towards a modification of the generalized diffeomorphism 
algebra in order to incorporate the $\star$-product. Therefore consistency 
conditions expected from diffeomorphism invariance need further investigation 
in 
these cases.

In what follows we illustrate some of the issues discussed above in an 
explicit example for the torus case.

 \subsection{$SU(3)$ example}
 \label{subsec:$SU(3)$ example}
We consider  the 2-torus compactification case in order to provide a 
specific example of the issues presented above.   The 
generalized momentum encoding KK and winding modes is $\check {\mathbb 
P}=(P^I,p_1,p_2; \tilde p^1,\tilde p^2)$. At a generic moduli point 
$\Phi=(g,b,A)$ non zero momenta lead to massive states. The massless vectors  
arise from zero modes $A_{\mu}^{\hat I(0)} \equiv 
A_{\mu}^{I(0)}, A_{\mu}^{1(0)}, A_{\mu}^{2(0)}$  and lead 
to the generic group  $U(1)_R^2\times U(1)^{16}$ gauge group. 
Enhancements will occur at specific moduli. As an example, 
let us look at moduli point $\Phi\equiv(g,b,0) $ with 
turned off Wilson lines. The set  of momenta that would 
lead to massless states at 
this point-recall \eqref{enhancementsector}-is $\check 
G(\Phi)_{480}=\{\check{\mathbb{P}}\equiv (\alpha ;0,0;0,0) \}$ where 
$\alpha \equiv (\underline {\pm1,\pm1,0,\dots})$ are the $SO(32)$ roots. Thus, 
when sliding to this moduli point these sates become massless and, together 
with the Cartan vectors should lead to  an enhancement to $U(1)_R^2\times 
SO(32)$ gauge group. Actually, we see from \eqref{structureconstant} 
that $f_{\mathbb{L} 
\mathbb {-L}  I}\equiv f_{\alpha -\alpha I}=\alpha^I$ providing the right 
structure constants involving charged fields and Cartans. 
Moreover, the phases arising from the $\star$-product-see for instance   
\eqref{modesfmunu} for the  field strength modes-now read 
from \eqref{chargessc} and  \eqref{so32phases}
\begin{equation}
 \tilde f_{\mathbb{L} \,
\mathbb {L}_2
\mathbb{L}_3}\equiv \tilde f_{\alpha \alpha_2 \alpha-\alpha_2}=-e^{i\frac12 
P_1EP_2}
\end{equation}
where $\frac12 E_{IJ}=\frac12(G_{IJ}+B_{IJ})$ is the $SO(32)$ Cartan-Weyl 
metric for $I\ge J$ and zero otherwise. These values exactly correspond to 
structure constants involving three charged fields (see for instance 
\cite{Green:1987sp}). 
We conclude that $S_{eff}(\Phi\equiv(g,b,0))$ corresponds to a well defined 
$U(1)_R^2\times SO(32)$ gauge theory.

Moduli points $\Phi\equiv(g,b,0) $ can lead to further enhancements  for 
specific values of $g$ and $b$ on the compactification 2-torus. It proves 
convenient to rewrite this point 
as $\Phi=(T,U,0)$ where 
$U=U_1+iU_{2}$ and $T=T_1+iT_{2}$ are the complex and the K\"{a}hler 
 structure of the torus respectively. They are  defined in terms of the metric 
and $b$  field as 
 $U_{1}  =\frac{g_{12}}{{g_{22}}},\, U_{2} =
\frac{\sqrt{\det g}}{{g_{22}}}, \, 2 T_{1} =  b_{12},\, 2 T_{2} =\sqrt{\det g}
$. An $SU(3)_L$ gauge symmetry enhancement occurs at point 
$\Phi_0=(-\frac12+i\frac{\sqrt3}2,-\frac12+i\frac{\sqrt3}2,0)$ which 
is  obtained from  the  $SU(3)$ Cartan matrix and $b$ field
 \be \label{su3cartan}
g_{mn}=\begin{pmatrix} 2 & -1 \\ -1 & 2 
\end{pmatrix}  
\qquad
b_{mn}=\begin{pmatrix} 0 & -1 \\ 1 & 0 
\end{pmatrix},
\ee
whereas at 
$\Phi_1=(i,i,0)$, associated to 
\be \label{su2cartan}
g_{mn}=\begin{pmatrix} 2 & 0 \\ 0 & 2 
\end{pmatrix} 
\ee
and $b=0$, an enhancement to 
$(SU(2)\times SU(2))_L$  group occurs.

At the $SU(3)$ point (some  basic information and notation is presented 
in Appendix \ref{sec:su3}), Left and Right momenta \eqref{leftrightmomenta} 
become 
\begin{eqnarray}
 l_{L}^{1} &=&\frac13(2p_1+p_2+{\tilde p}^{1}+{\tilde p}^{2}), \qquad
 l_{L}^{2} =\frac13(p_1+2p_2-{\tilde p}^{1}+2{\tilde 
p}^{2}),\\
 l_{R}^{1} &=&\frac13(2p_1+p_2-2{\tilde p}^{1}+{\tilde p}^{2}), \qquad
l_{R}^{2} =\frac13(p_1+2 p_2-{\tilde p}^{1}-{\tilde p}^{2}),
\end{eqnarray}
whereas at $SU(2)\times SU(2)$ we have
\begin{eqnarray}
 l_{L}^{1} &=&\frac12(p_1+{\tilde p}^{1}), \qquad
 l_{L}^{2} =\frac12(p_2+{\tilde 
p}^{2}),\\
 l_{R}^{1} &=&\frac12(p_1-{\tilde p}^{1}), \qquad
l_{R}^{2} = \frac12(p_2-{\tilde p}^{2}),
\end{eqnarray}
where we have set $\alpha '=1$. 

It is easy to check that the six weights 
\begin{equation}
\check {\mathbb 
P}_0= \pm (0;0,1; 1,1),  \pm (0;1,0; 1,0),\pm (0;-1,1;0,1)               
\end{equation}
lead to 
$l_R^m=0$ 
 and  
to $(l_{L}^1, l_{L}^2)=\pm(1,1), \pm(1,0),\pm(0,1)$. The latter are the  
coordinates\footnote{Recall that $l_{L}^m$ are 
coordinates of the 
weight vectors of the representation in  the simple root lattice, namely 
$\Lambda= l_{L}^m\alpha_m$ with $\alpha_m$ the simple roots, whereas 
$l_{Lm}=g_{mn} l_{L}^m$ correspond to coordinates (Dynkin labels) in the 
dual 
(weight) lattice.} in the simple root lattice base  $\alpha_1=(\sqrt{2};0)$,
$\alpha_2=(-1/\sqrt{2};\sqrt{3/2})$ and satisfy  (see 
\eqref{su3cartan})
$l_L^2=l_L^m g_{mn}l_L^m=2 \tilde p^m p_m=2$. They 
label the  six massless charged vectors of 
$SU(3)$.  We collected this information in  Table  
\ref{table:su3rep8}. 
In particular 
$(1,1)$ corresponds to the highest weight of the adjoint $\bf 8$ 
representation. Together with  the $SO(32)$ modes above the set 
$\check 
G(\Phi_0)_{480+6}=\{(\alpha ;0,0;0,0); \pm 
(0;0,1; 1,1),  \pm (0;1,0; 1,0),\pm (0;-1,1;0,1)\}$  defines, by similar 
considerations as above,  a heterotic low energy effective theory  
$S_{eff}(\Phi_0)$ with gauge group $U(1)_R^2\times SU(3)\times SO(32)$.
 
On 
the other 
hand, the generalized momenta $\check {\mathbb 
P}_1=\pm (0;1,0; 1,0),\pm (0;0,1;0,1)$ provide the charged massless vectors of 
$SU(2)\times SU(2)$ at $\Phi_1$. We notice that states (vectors and 
scalars) associated to  modes  $\pm 
(0;1,0, 1,0)$ are massless at both points whereas for $\pm (0;0,1;0,1)$ we 
have, 
at $ \Phi_0$ point, 
$l_{L} =\pm(\frac23, \frac43), \,l_{R}=\pm(\frac23 \pm \frac13)$ that satisfy
$l_{L}^2-l_{R}^2=2$ and correspond to very massive states with 
$\ap m^2=\frac43$. That is why it drops out from the 
effective low energy theory at $ \Phi_0$ point. 
Thus, the low energy theory at $\Phi_1$ corresponds to 
$S_{eff}(\Phi_1)$ with gauge group $U(1)_R^2\times SU(2)\times SU(2)\times 
SO(32)$ where the charged vectors  arise  from $\check {\mathbb 
P}_1$ modes together with $SO(32)$ modes above, namely from states in 
$\check G(\Phi_1)_{480+4}=\{(\alpha ;0,0;0,0); \pm 
(0;0,1; 1,1),  \pm (0;1,0; 1,0),\pm (0;-1,1;0,1)\}$. 

Having discussed  the low energy effective action arising from action 
\eqref{hetactionuplift} at different moduli points,  we propose 
to explore the contributions from massive states. 
Let us concentrate in the $\Phi_0$ point.  

For instance, the massive state $l_{L} 
=(-\frac23, -\frac43)$,  not contributing to  the low energy theory,
must  now be considered. Interestingly enough, this state corresponds to 
the lowest weight of the symmetric 
${\bf 6}$ representation (whereas $ 
(\frac23, \frac43)$ corresponds to the highest weight of ${\bf\bar 6}$). 
Indeed,  it is easy to see that  when 
shifting with  $\check {\mathbb P}_0$ vectors we obtain   the mode vectors 
$(\Lambda_{2},q_R) $ with 
$\Lambda_{s}=({l_L}^1_s,{l_L}^2_s)$, 
$q_R=(-\frac23,-\frac13)$, with $s=1,\dots 6$,  where
\begin{equation}
     \Lambda_{s}\equiv 
(\frac43,\frac23),(\frac13,\frac23),(-\frac23,\frac23),(\frac13,-\frac13),
(-\frac23,-\frac13) ,(-\frac23,- \frac43).                                      
  \end{equation}
These modes fill the  ${\bf 6}_{(-\frac23,-\frac13) }$ representation  of 
$U_R(1)^2\times SU(3)$ (and similarly  ${\bf\bar 6}_{(\frac23,\frac13) }$). 

We notice that all states have  $l_R^2= \frac23$ but $l_L^2= \frac83$ for 
$s=1,3,6$ whereas  $l_L^2= \frac23$ for $s=2,4,5$. Indeed, they satisfy 
\begin{equation}
 \frac12 l_{Ls}^2-\frac12 l_{Rs}^2=\frac12l_{Ls}^2- \frac13=\tilde 
p_s.p_s=1-N_s
\end{equation}
with $N_s=0$ for $s=1,3,6$ and $N_s=1$ for $s=2,4,5$. Moreover, for these 
values of $N_s$, all states have the same mass $\ap m^2=\frac43$, as it must 
be if they all belong to the same multiplet. 
Thus we conclude that states with oscillator numbers $N=1,0$ must be mixed in 
order to build up the $\bf 6$ massive symmetric representation.
These results are collected in Table \ref{table:su3rep6}.
\begin{table}[ht]
 { \scriptsize{ 
 \begin{minipage}{.5\linewidth}
      \centering
    \vspace{2,5 cm}    \begin{tabular}{|c|c|c|c| }
 \hline
& & &\\[-0.4cm]
 Dynkin label& $(p_{1},p_{2};\tilde p^{1},\tilde p^{2}) $ & 
$(l_{L}^{1},l_{L}^{2})$ & $N$\\
\hline
  & & &\\
$\pm  (1,1)$  &$ \pm(0,1;1,1)$ &$ \pm(1,1)\equiv\pm \alpha_3$& $0$\\ \hline
& &  & \\
$ \pm (2,-1)$ & $  \pm(1,0,;1,0)$  & $\pm (1,0)\equiv \pm\alpha_1$ &  $0$\\ 
\hline
& &  & \\
$\pm  (-1,2)$ & $\pm(-1,1;0,1)$  & $\pm (0,1)\equiv \pm\alpha_2$ & $0$  \\ 
\hline
 & &  & \\
$2\times (0,0)$ & $(0,0;0,0)$  & $ (0,0)$ &1  \\ 
\hline
\end{tabular}
\caption{${\bf 8}_{(0,0)}$ }
\label{table:su3rep8}
\end{minipage}   
 \begin{minipage}{.45\linewidth}
      \centering
      \begin{tabular}{|c|c|c|c| }
 \hline
& & &\\[-0.4cm]
r$ \equiv $Dynkin label& $(p_{1},p_{2};\tilde p^{1},\tilde p^{2}) $ & 
$(l_{L}^{1},l_{L}^{2})$ & $N$\\
\hline
  & & &\\
$1\equiv (2,0)$  &$ (0,1;2,1)$ &$ (\frac43,\frac23)$& $0$\\ \hline
& &  & \\
$2\equiv (0,1)$ &   (-1,1,;1,1)  & $ (\frac13,\frac23)$ &  $1$\\ \hline
& &  & \\
$3\equiv (-2,2)$ & $(-2,1;0,1)$  & $ (-\frac23,\frac23)$ & 
$0$  \\ \hline
 & &  & \\
$4 \equiv (1,-1)$ & $(0,0;1,0)$  & $ (\frac13,-\frac13)$ &1  \\ \hline
& &  & \\
$5 \equiv (-1,0)$ & (-1,0;0,0)  & $ (-\frac23,-\frac13)$ &1 \\ \hline
& &  & \\
$6  \equiv (0,-2)$ & (0,-1;0,-1)  & $ (-\frac23,-\frac43)$ &0 \\ \hline
\end{tabular}
\caption{${\bf 6}_{(\frac23,\frac13) }$ representation. }
\label{table:su3rep6}
\end{minipage} \hspace{1cm}
\caption*{{ \footnotesize{ Some relevant data for ${\bf 6}_{(\frac23,\frac13) 
}$ symmetric and ${\bf 8}_{(0,0)}$ adjoint representations of $U(1)^2\times 
  SU(3)$ is provided. In the first column a number is associated to each pair 
of 
Dynkin  coordinates in weight 
space. The second column presents their corresponding KK momenta and windings 
  whereas in the third column the weights are given in the root basis. The last 
column 
  indicates the oscillator number required by level matching.}}}}}
  \label{table:}\nn
\end{table}


 These states, even though they are very massive,  are indeed  present in our 
construction. 
As an illustration let us consider the massive  scalar 
fields with mass $\ap m^2=\frac43$. The $N=0$ states  
correspond to the modes $M_{\bar J}(x)^{( \Lambda_{s},q_R)}$  with $r=1,3,6$ in 
the GKK expansion of $M_{\bar J}(x,{\mathbb Y})$ whereas states 
with $N=1$,  $M_{\bar J m}(x)^{( \Lambda_{s},q_R)}$, with 
$s=2,4,5$,
are contained in  $M_{\bar J m}(x,{\mathbb Y})$ expansion. It is worth 
noticing that in the $N=1$ case there are two states $m=1,2$ for each mode  
$\Lambda_{s}$ so we expect that only a combination of them enters to complete 
the representation. We discuss this issue below 
\footnote{In what follows, in order to lighten the notation,  we  avoid 
indicating the charge 
$q_R$ of each state, being the same for all states in the multiplet.}. 

It is enlightening to look at the ``covariant'' derivative 
  for the modes  $\Lambda_{s}$ above. For $N=0$ modes, $s=1,3,6$,  the 
expression \eqref{covderivativemode} must be 
considered. For these states it can be expressed as
\begin{eqnarray}
\label{covderivativemode6repn0}
  \mathcal{D}_{\mu}M^{(\Lambda_{s})}_{\bar 
J}&=&\partial_{\mu}M^{(\Lambda_{s})}_{\bar J}+
ig\sum_{l}'\tilde{f}_{\Lambda_{s}\alpha_l{\Lambda_{s}-\alpha_l}}A_{\mu}^{ 
(\alpha_l)}M^{(\Lambda_{s} -\alpha_l)}_{ \bar J}
 \\\nn
&+& ig\sum_{l}'
\tilde{f}_{\Lambda_{s}\alpha_l{\Lambda_{s}-\alpha_l}} 
A_{\mu}^{ 
(\alpha_l)}\alpha_l^m M_{m\bar J}^{(\Lambda_{s} -\alpha_l)}
+
ig\Lambda_{sm}A_{\mu}^{m(0)} 
M^{(\Lambda_{s})}_{\bar J}
+igq_{R\bar{I}} 
A_{\mu}^{\bar{I} (0)}M^{(\Lambda_{s})}_{\bar J}+...
 \end{eqnarray}
where we have just shown the terms that couple to $U_R(1)^2\times SU(3)$ 
gauge vectors, the \dots encoding all the 
rest. 
On the other hand, for $N=1$ states, $s=2,4,5$, the derivative 
\eqref{covderivativeKKscalarmode} reads 
\begin{eqnarray}
\label{covderivativemode6repn1}
  \mathcal{D}_{\mu}M^{(\Lambda_{s})}_{m \bar 
J}&=&\partial_{\mu}M^{(\Lambda_{s})}_{m\bar J}+
ig\sum_{l}'\tilde{f}_{\Lambda_{s}\alpha_l{\Lambda_{s}-\alpha_l}}A_{\mu}^{ 
(\alpha_l)}\alpha_l^mM^{(\Lambda_{s} -\alpha_l)}_{ \bar J}
\\\nn
&+&
2ig\sum_{l}'
\tilde{f}_{\Lambda_{s}\alpha_l{\Lambda_{s}-\alpha_l}} 
A_{\mu}^{ 
(\alpha_l)} M_{m \bar J}^{(\Lambda_{s} -\alpha_l)}+ ig{\Lambda_{s}}_{k} 
A_{\mu}^{ k (0)}M^{(\Lambda_{s})}_{m \bar J}         +   
igq_{R\bar{I}} 
A_{\mu}^{\bar{I}(0)}M^{(\Lambda_{s})}_{m\bar J}+...
 \end{eqnarray}
 
The last term is just the expected coupling to $U(1)_R^2$ vectors.
The $\alpha_l$ index labels the $6$ charged vectors of $SU(3)$ in 
correspondence with $SU(3)$ roots where 
$\alpha_4=-\alpha_1,\alpha_5=-\alpha_2,\alpha_6=-\alpha_3$ (see Table
\ref{table:su3rep8}). $A_{\mu}^{m(0)}$ are the Cartan gauge vector fields.

We observe that in the first equation \eqref{covderivativemode6repn0} linear 
combinations of $N=1$ modes $\alpha_l^m M_{m \bar J}^{(\Lambda_{s} -\alpha_l)}$ 
do appear. Indeed, as mentioned above, we expect linear  combinations to 
provide the physical degrees of freedom entering in the $\bf 6$ multiplet.

Before presenting some explicit examples recall that  a well defined 
covariant derivative on fields $\Phi^{s}$ in this multiplet must read 
\begin{eqnarray}
\label{covderivative6rep}
  \mathcal{D}_{\mu}\Phi^{s}&=&\partial_{\mu}\Phi^{s}+ ig 
(T_{\alpha_l})_{sr} A_{\mu}^{ 
(\alpha_l)}\Phi^{r}+ig 
(T_I)_{sr} A_{\mu}^{I}\Phi^{r}
 \end{eqnarray}
where $T_{\alpha_l}$ and  $T_I$  are  the matrices 
corresponding to $SU(3)$  charged and Cartan generators, respectively,  in the 
$\bf 6$ representation. They are 
collected (in Cartan-Weyl basis)  in Appendix \ref{sec:su3}.

Let us consider the derivative \eqref{covderivativemode6repn0} for the 
state $M^{(\Lambda_{1})}_{\bar 
J}$. It reads
\begin{eqnarray}\nn
\label{covderivative1}
  \mathcal{D}_{\mu}M^{(\Lambda_{1})}_{\bar 
J}&=&\partial_{\mu}M^{(\Lambda_{1})}_{\bar J}+
ig\tilde{f}_{1\alpha_3 4}A_{\mu}^{ 
(\alpha_3)}\alpha_{3}^mM^{(\Lambda_4)}_{ i\bar J}+
ig\tilde{f}_{1\alpha_1 2}A_{\mu}^{ 
(\alpha_1)}\alpha_{1}^mM^{(\Lambda_2)}_{i \bar J}\\
&+&
ig(\sqrt{2} A_{\mu}^{1(0)}+ \sqrt{\frac2 3}A_{\mu}^{2(0)})
M^{(\Lambda_{1})}_{\bar J}
 \end{eqnarray}
where we have used that $\Lambda_{1}=\frac43\alpha_1+\frac23\alpha_2
=(\sqrt{2},\sqrt{\frac2 3})$. Also we denote 
$\tilde{f}_{\Lambda_s\alpha_j\Lambda_r}=\tilde{f}_{s\alpha_j r}$. By using that 
$\tilde{f}_{1\alpha_1 
2}=\tilde{f}_{1\alpha_3 4}=1$  and by defining
\begin{eqnarray}\label{normalizations1}
\Phi_{\bar J}^1=M^{(\Lambda_{1})}_{\bar 
J},\qquad \Phi_{\bar J}^2=\frac1{\sqrt2} \alpha_{1}^mM^{(\Lambda_2)}_{m \bar 
J}, \qquad\Phi_{\bar J}^4=-\frac1{\sqrt2}\alpha_{3}^mM^{(\Lambda_4)}_{m 
\bar 
J}
\end{eqnarray}
this derivative can be recast in the form \eqref{covderivative6rep} with  
$(T_{\alpha_1})_{12}=(T_{\alpha_3})_{14}=\sqrt2$ and $({T_1})_{11}=\sqrt{2}, 
({T_2})_{11}=\sqrt{\frac2 3}$ in exact correspondence with 
\eqref{T6}.
Nevertheless, we should check that these definitions are consistent for the six 
states. Consider, for instance,  the derivative of  $ M^{(\Lambda_{2})}_{i \bar 
J}$ field. We have noticed above that this field appears contracted 
with root $\alpha_1$ in order to define the field $\Phi_{\bar J}^2$ with the 
correct transformation properties. This indicates that  we must actually 
compute the derivative of $\Phi_{\bar J}^2$. Thus, by projecting in 
\eqref{covderivativeKKscalarmode} we find 
\begin{eqnarray}\nn
\label{covderivativephi2}
  \mathcal{D}_{\mu}\Phi_{\bar J}^2&=&\partial_{\mu}\Phi_{\bar 
J}^2+
ig\frac1{\sqrt2}\tilde{f}_{2 (-\alpha_1) 1}A_
{\mu}^{ 
(-\alpha_1)}(-\alpha_1^2)M^{(\Lambda_{1})}_{ \bar J}
+ig
\tilde{f}_{2 \alpha_3 5} 
A_{\mu}^{ 
(\alpha_3 )}\frac2{\sqrt2}\alpha_1^m M_{m \bar J}^{(5)}\\\nn
&+& ig
\tilde{f}_{2 \alpha_2 4} 
A_{\mu}^{ 
(\alpha_2)}\frac2{\sqrt2}\alpha_1^m M_{m \bar J}^{(4)}+ig
\tilde{f}_{2 \alpha_1 3} 
A_{\mu}^{ 
(\alpha_1)}\frac2{\sqrt2}\alpha_1^m M_{m \bar J}^{(3)}+ 
ig{\sqrt{\frac23}
A_{\mu}^{2 (0)}\Phi_{\bar J}^2} \\
 \end{eqnarray}
Interestingly enough we see that $({T_1})_{22}=0,({T_2})_{22}=\sqrt{\frac23}$ 
as expected from \eqref{T6}. Also,   since $\tilde{f}_{2 
(-\alpha_1) 1}= -\tilde{f}_{2 \alpha_3 5}=\tilde{f}_{2 \alpha_2 4}= \tilde{f}_{2 
\alpha_1 3}=1$, 
consistency requires  the extra definitions  
\begin{eqnarray}\label{normalizations2}
\Phi_{\bar J}^3=M^{(\Lambda_{3})}_{\bar 
J},\qquad \Phi_{\bar J}^4=-\frac2{\sqrt2} \alpha_{1}^mM^{(\Lambda_4)}_{m 
\bar 
J}, \qquad\Phi_{\bar J}^5=\frac2{\sqrt2}\alpha_{1}^mM^{(\Lambda_5)}_{m 
\bar 
J}
\end{eqnarray}
in order to have $(T_{-\alpha_1})_{21}=\sqrt2$ and 
$(T_{\alpha_2})_{24}=(T_{\alpha_3})_{25}=1$ (see  \eqref{T6}).
However, we have already defined $\Phi_{\bar J}^4$ in \eqref{normalizations1},
so the only way to obtain a consistent description is to have
\begin{equation}\label{m4projection}
 (2\alpha_1-\alpha_{3})^mM^{(\Lambda_4)}_{m\bar J}= 
(\alpha_1-\alpha_{2})^mM^{(\Lambda_4)}_{m\bar 
J}=\Lambda_4^mM^{(\Lambda_4)}_{m\bar J}=0.
\end{equation}
Therefore, we observe  that the two fields ($m=1,2$)  
$M^{(\Lambda_4)}_{m\bar 
J}$ must combine into the physical state $\alpha_{3}^mM^{(\Lambda_4)}_{m\bar 
J}$ and the orthogonal state $(\alpha_1-\alpha_{2})^mM^{(\Lambda_4)}_{m\bar 
J}$ (notice that $\alpha_{3}(\alpha_1-\alpha_{2})=0$) that must decouple.
Indeed, by completing the computation of the derivatives for the rest of the 
states it can be checked that  complete   derivative \eqref{covderivative6rep} 
with $T^a$ given in \eqref{T6} is reproduced if  the condition
\begin{equation}
 \Lambda_{s}^mM^{(\Lambda_s)}_{m\bar J}=0
 \label{scalarphysicalstates}
\end{equation}
is imposed for $s=2,4,5$. The normalized fields are defined as  
\begin{eqnarray}\label{normalizatiosscalars}
\Phi_{\bar J}^s &=& M^{(\Lambda_{s})}_{\bar 
J},\qquad s=1,3,6\\\nn
\Phi_{\bar J}^2&=&\frac1{\sqrt2} 
\alpha_{1}^mM^{(\Lambda_2)}_{m \bar 
J}, \qquad\Phi_{\bar J}^4=-\frac1{\sqrt2}\alpha_{3}^mM^{(\Lambda_4)}_{m 
\bar 
J},\qquad\Phi_{\bar J}^5=-\frac1{\sqrt2}\alpha_{2}^mM^{(\Lambda_5)}_{m 
\bar 
J}.
\label{physicscalars}
\end{eqnarray}
The $\alpha_l$ appearing in the contraction with $M^{(\Lambda_s)}_{m 
\bar 
J}$ satisfy $\alpha_l.\Lambda_s=0$. Namely, they select the DOF of 
$M^{(\Lambda_s)}_{m 
\bar J}$ orthogonal to  $\Lambda_s$ as the physical one.

Since massive vector bosons have the same weights as the scalars, the same line 
of reasoning leads to a consistent covariant derivative 
of the massive vector fields, from expressions \eqref{modesfmunu} and 
\eqref{cartanfs}.  The 
physical vector bosons are obtained from \eqref{physicscalars} just by 
replacing $M^{(\Lambda_{s})}_{\bar J}\rightarrow A^{(\Lambda_{s})}_{\mu}$ and 
$M^{(\Lambda_{s})}_{m\bar J}\rightarrow A^{m(\Lambda_{s})}_{\mu}$. Namely,
$  A^{s}_{\mu}= A^{(\Lambda_{s})}_{\mu}$ for $s=1,3,6$ and  $ 
A^{s}_{\mu}=\frac1{\sqrt2} \alpha_l^m A^{(\Lambda_{s})}_{m\mu}$ for $s=2,4,5$ where   $\alpha_l.\Lambda_{s}=0$. The  physical 
vectors  satisfy
\begin{equation}
 \Lambda_{s}^m A^{ (\Lambda_s)}_{m\mu}=0.
 \label{vectphysicalstates}
\end{equation}

The physical degrees of freedom (DOF)  conditions above  can be interpreted 
from different perspectives. From a string theory point of view this 
requirement arises from conformal invariance. Namely, by looking at the OPE of 
the stress energy tensor with  the vertex operators associated to different 
$N=1$ modes above ($r=2,4,5$) an anomalous  term 
$\frac1{z^3}{\Lambda_{s}^mM^{(\Lambda_s)}_{m\bar J}}$ is generated for scalars 
and $\frac1{z^3}{\Lambda_{s}^m A^{(\Lambda_s)}_{m \mu }}$ for vectors. Absence 
of anomalies leads to the physical state condition. 
Even if our  whole 
construction emerges from string theory we would like to deal with consistency 
conditions contained in the proposed action without introducing external 
information. 

The DOF conditions can also be interpreted as inherited from 
consistency in 10 dimensions. Schematically,   $N=1$ modes can be interpreted 
 as KK reductions of a generalized metric $ \cal H^{\cal M \cal N}$ with ${\cal 
M, 
\cal N}=1,\dots 10$, encoding the massless metric $g_{\cal M \cal N}$ and 
anti-symmetric field  
$b_{\cal M  \cal N}$, satisfying the  gauge condition 
$\partial_{\cal M}{\cal H}^{\cal M \cal N}=0$. By splitting 
indices into space time and compactification indices, namely ${\cal M}\equiv 
\{\mu,m\}$,  this condition gets  splitted as 
\begin{eqnarray}
 \partial_{\mu}{\cal H}^{\mu \nu}(x,{\mathbb Y})+ \partial_{m}{\cal H}^{m 
\nu}(x,{\mathbb Y}) &=&0\rightarrow \partial_{\mu}{\cal H}^{\mu 
\nu}(x)^{(\mathbb L)}+ {\mathbb L}_{m}{\cal H}^{m \nu}(x)^{(\mathbb 
L)}=0\\
  \partial_{\mu}{\cal H}^{\mu n}(x,{\mathbb Y})+ \partial_{m}{\cal H}^{m 
n}(x,{\mathbb Y})&=&0\rightarrow  \partial_{\mu}{\cal H}^{\mu n}(x)^{(\mathbb 
L)}+
 {\mathbb L}_{m}{\cal H}^{m 
n}(x)^{(\mathbb 
L)}=0.
\end{eqnarray}
For massless modes, corresponding to ${\mathbb L}=0$, we recover the expected 
gauge conditions for the $g_{\mu \nu}$, the B field contained in ${\cal H}^{\mu 
\nu}(x)^{(0)} $ and the gauge vectors 
 $A_{m \mu} \equiv {\cal H}^{(0)}_{\mu n}(x)$. 
 However, for massive modes, a consistency requirements  $\partial_{\mu}{\cal 
H}^{\mu n}(x)^{(\mathbb 
L)}=0$ is needed for a Proca field to have the right number of degrees of 
freedom. Similarly  $ \partial_{\mu}{\cal H}^{\mu 
\nu}(x)^{(\mathbb L)}=0$ for massive symmetric and anti-symmetric tensors. 
We conclude that  
\begin{equation}
 {\mathbb L}_{m}{\cal H}^{m 
n}(x)^{(\mathbb 
L)}={\mathbb L}_{m}{\cal H}^{m \nu}(x)^{(\mathbb 
L)}=0
\end{equation}
which correspond to physical state conditions 
\eqref{scalarphysicalstates} and \eqref{vectphysicalstates} respectively.

Finally let us provide a third way of looking at physical DOF conditions. 
Given the two $N=1$ fields, $m=1,2$ we have seen that we can combine them into 
two (orthogonal) linear independent combinations. For instance,  in the case of 
$M^{(\Lambda_4)}_{m\bar 
J}$ fields above \eqref{m4projection}, we can consider the l.i. combinations 
$(\alpha_1-\alpha_{2})^mM^{(\Lambda_4)}_{m\bar 
J}= \Lambda_4^mM^{(\Lambda_4)}_{m\bar 
J}$ and $(\alpha_1+\alpha_{2})^mM^{(\Lambda_4)}_{m\bar 
J}$. The second one provides the 
$\Phi^4_{\bar J}$ physical DOF whereas the first term should not be present in 
the spectrum. Similarly, the  vector boson  combination   $\Lambda_{4}^m A^{ 
(\Lambda_4)}_{m\mu}$ must decouple. The same reasoning holds for $s=2,4,5$. 

Now, recall that the first 
two rows in \eqref{hetactionsplit} contain massive field modes 
corresponding to graviton and Kalb-Ramond $g_{\mu\nu}^{(\mathbb L)}, 
B_{\mu\nu}^{(\mathbb L)}$. These modes  satisfy ${\mathbb 
L^2}=l_l^2-l_R^2=0$ level matching condition. It is easy to see 
that the lowest massive levels have mass $\alpha' m^2=\frac43$  and 
correspond to weights $\pm \Lambda_s$ with 
$s=2,4,5$ and $U(1)_R^2$ charges $q_R=\pm(-\frac23,-\frac13)$  as the vector 
and scalars states 
discussed above.

With this observation in mind, the decoupling  can be understood (see 
\cite{amn} for a related discussion in a DFT context) as follows: The scalars $ 
\Lambda_s^mM^{(\Lambda_s)}_{m\bar 
J}= \Lambda_s\cdot M^{(\Lambda_s)}_{\bar 
J}   $ are ``eaten'' by the corresponding vector boson to become massive. At 
the same time a massive graviton mode $g_{\mu\nu}^{(\Lambda_s)}$ ``eats'' the 
vector boson to become a massive graviton with the correct degrees of freedom.
Indeed, this can be explicitly shown by following the construction in 
Ref.\cite{amn}. Namely,   we can write write the massive vector meson as 
\begin{equation}
 A^{'(\Lambda_s)}_{m\mu}= A^{(\Lambda_s)}_{\mu}-\frac1{m^2}\partial_{\mu
}
(
\Lambda_s\cdot M^{ (\Lambda_s)}_{ \bar 
J}M^{(\Lambda_s) \bar 
J}_m)
\end{equation} such that the $\Lambda_s\cdot A^{'(\Lambda_s)}_{\mu}$ projection 
is 
``eaten''  by the massive graviton 
\begin{eqnarray}
 g_{\mu\nu}^{s}&=& 
g_{\mu\nu}^{(\Lambda_s)}-\frac1{m^2}\Lambda_s\cdot 
\partial_{(\nu}A^{'(\Lambda_s)}_{\mu)}
\end{eqnarray}
and the remaining physical states 
$A^{s}\propto \alpha_l. A^{'(\Lambda_s)}_{m\mu}$, with 
$\alpha_l\cdot \Lambda_s=0$  
satisfy  $\partial_ {\mu}  A^{'\mu s}=0$.

Moreover, by noticing that the   weights in the fundamental  
representation  of $SU(3)$ correspond to modes 
$\Lambda_s\equiv(0,1),(1,-1),(0,-1)$,  
 we see that gravitons organize into multiplets of
$(\bf 
3)_{(-\frac23,-\frac13)}$ (and $\bf {\bar 3}(\frac23,\frac13)$)  of the gauge 
group  $U(1)_R^2\times 
SU(3)$. 

Let us close this section by stressing that the action 
\eqref{hetactionuplift} appears to contain very non trivial information even 
for states with masses of the order of the string mass. This is what the 
analysis of the covariant derivative of the massive symmetric representation 
in the above example indicates.
Again, going to higher massive states  would require the introduction of 
$N >1$ and call for further investigation.

\section{Summary and Outlook}
\label{sec:Summary and Outlook}
A striking and distinctive feature of string compactifications is that, at 
certain values of the compactification background -namely a point in moduli 
space-  compact momenta and winding modes   can combine to generate new (let 
us  say $n_c$)  
massless vector bosons   leading to an enhancement of the  gauge symmetry 
group. 
Different enhancement can occur for other values of moduli and, generically, 
for other values of winding and 
momenta. In the notation presented above 
 \eqref{enhancementsector} we would say  that, for a given number of 
compact 
dimensions $r$,   several sets  $  \check 
G(\Phi_i)_{n^i_c}$ of generalized 
momenta $\check {\mathbb P}$  could exist. These  lead to enhancement at 
moduli point $\Phi_i$ where $n^i_c$ vector bosons and scalars become massless. 
The structure 
gets richer for lower space-time dimensions.

We have shown that the heterotic low energy effective theory at each 
$\Phi_i$ is obtained by  considering fields associated to 
$\check {\mathbb P} \in  \check 
G(\Phi_i)_{n^i_c}$ modes and the zero modes arising from fields in the gravity 
sector whereas all other, very massive modes, are integrated out.
Slight displacements  from $\Phi_i$ can be interpreted as a Higgs 
mechanism. 
Actually, when  moving along moduli space  some (or all) of these fields 
become massive whereas other fields become lighter at a different point. 
Therefore,  a moduli dependent description able to account for these 
different 
enhancements implies handling an infinite number of fields.
In this work we were able to identify some guiding lines towards this 
description, which is 
encoded in a moduli dependent effective 
action where a non-commutative $\star$-product plays a central role.

The proposed action, written in $d$ space time dimensions, contains a 
generically infinite number of fields labeled by allowed momenta and winding 
modes. In principle, this action  could have been obtained by carefully 
looking at string 3-point amplitudes  of vertex operators associated with 
these modes. 
We have shown that these infinite fields in $d$ dimensions can be 
understood as modes of a GKK expansion in the internal double torus   and 
heterotic coordinates $\mathbb Y\equiv (y^I, y_L^m, y_R^m )$, 
providing an uplifting to higher dimensions. 
In this sense the action can be seen as a Kaluza-Klein inspired rewriting of 
a double field theory (see for instance Ref.\cite{Hohm:2013nja}), where 
coordinates are split into space-time coordinates (that could be formally 
doubled)  and internal double coordinates. However, once compact coordinates 
come into play we   noticed that a $\star $-product that introduces 
a non-commutativity in the target compact space is called for.
Indeed, it is this non-commutativity that leads to the adequate 
factors to reproduce the structure constants. As we have shown in an example, 
this non-commutativity  also appears to have the right 
features to reproduce  the  generator 
matrix elements in higher order  representations where massive states live,  as 
required by  gauge invariance. 
It would be interesting to trace  the origin of this product for the 
heterotic string case \cite{mp} back. In the context of bosonic string it was 
shown  
in 
\cite{leigh} to be associated to non-commutativity of string coordinate zero 
modes.

An interesting result of the construction is that, close to a given enhancement 
point $\Phi_0$, by keeping just the $n_c$ slightly massive fields, the Higgs 
mechanism can be cast in terms of $\tilde f$ moduli dependent ``structure like
constants'' that become the enhanced group  structure constants at $\Phi_0$. 
This description provides a field theory stringy version of the gauge symmetry 
breaking-enhancement mechanism. This fact  was already addressed in 
the 
context of DFT in \cite{aamr,aamp} where  it was shown that 
constants $\tilde f(\Phi)$ can be interpreted 
as DFT Scherk -Schwarz \cite{reviewamn, effectiveSS,edft} compactifications 
generalized fluxes. These fluxes can be read from the DFT generalized 
diffeomorphism algebra. Actually, it is worth noticing that 
these fluxes were explicitly 
constructed from a generalized frame  only in the circle case where a $SU(2)$ 
enhancing at the self dual radio $R=\tilde R=\sqrt{\ap}$ occurs
\cite{agimnr,aamr,Cagnacci:2017ulc, aamp}. 
Interestingly enough, the $SU(2)$ case is the only situation where the 
$\star$-product is not needed (essentially due to the absence of a $b$ 
field).   Difficulties in going beyond this 
case were mentioned in \cite{aamr,Cagnacci:2017ulc}. In   
Ref.\cite{Cagnacci:2017ulc,Fraiman} a connection among these 
difficulties and vertex operators  cocycle factors was suggested. The  
non-commutative  product could 
provide a solution for this problem since the $\star $ appears as  
a manifestation of the cocycle factors in the DFT context. 
Let us stress that the $\star$-product is 
not needed, 
at third order in fluctuations, if fields satisfy ${\mathbb L}^2=0$ level 
matching condition. This is why it did not manifest in original DFT 
constructions  but  would be 
required in a 
DFT formulation including four (or higher) order terms in the fields, 
where cocycle factors would be required, as 
it stressed in \cite{Hull:2009mi}.

Actually,  the problem already arises at third order
when the  Lie algebra of three charged fields  with $\frac12 
{\mathbb L}^2=1$ LMC is considered, which is just the situation where the 
$\star$-product  phase is relevant. Moreover,  since  $\star $-product 
is providing cocycle factors,
four order terms (or higher)  could be consistently considered in 
DFT. In fact, we have shown that this appears 
to be the case  in a partial computation of the fourth order scalar potential. 

As mentioned 
above,  a modified version of generalized diffeomorphisms is called for to 
handle 
 these cases. 
The detailed construction is left for future investigation.

 In our construction we started by proposing mode expansions restricted by the 
level matching constraint  $\frac12 \mathbb L^2=1$  (corresponding to  $N=0$ 
oscillators)  necessary to contain massless 
vectors at the enhancement point. Even if it effectively interpolates 
among different enhancement points, we stressed that new ingredients must be 
incorporated. In particular, at first mass level, we noticed that   for 
massive states to organize   into multiplets of the enhanced group $G$,  $N=1$
oscillator number is also required. Since we had already included the  $N=1$ 
case, to tackle the gravity sector, we showed in an example that indeed 
massive vector and scalar states nicely fill $G$ multiplets for first massive 
level. This happens to be the case also for gravity sector massive 
modes.
 However, for higher masses, other oscillator numbers are expected (this 
was  was noticed in \cite{amn}). Namely, if we consider next to first massive 
level, in order to complete  a $G$ multiplet, a level matching condition 
 with $N>1$ is required\footnote{These observations are based on the 2-torus example 
 of \ref{subsec:$SU(3)$ example} with Wilson lines turned off. The general situation with WL and/or higher dimensional torus 
 needs further investigation. }. 
 We see that a simple gauge symmetry 
consistency check  points towards the 
necessity of including massive higher spin fields and higher derivative terms 
in 
the action (see \cite{jjm}, 
 \cite{vdv} for a discussion from another perspective), as is in fact expected 
from 
string theory.
Let us stress that  gauge invariance underscores the 
limitations of 
the 
construction but at the same time it is  a guide for  consistent 
extensions. Indeed, gauge invariance provides a tool to  
systematically  include higher spin modes and $\ap$ corrections by looking for 
consistency all the way  from the very first massive levels  up 
to the highest ones.

Throughout our construction we have made intensive use of 
DFT tools. In 
particular, before mode expanding, all fields are expressed in terms of higher 
dimensional coordinates. However, a fully higher dimensional version is 
still  lacking in the sense that fields are written here in terms of space time 
$d$ dimensional indices.  Formally it appears rather straightforward. 
On the one hand,   the  new fields we are 
introducing here associated to $N=0$, can be cast in terms of  a $D$ 
dimensional ``charged vector''  field $A^{\cal M}(x,\mathbb 
Y) \equiv (A^{\mu}(x,\mathbb 
Y), M^{\cal I}(x,\mathbb 
Y)) $ and, on the other hand,  the sectors 
originating in the generalized metric in 10-dimensions  were  already 
addressed in \cite{amn} (up 
to a third order expansion) in terms of a generalized metric.  However, it 
appears that the latter must be modified by the presence of the new 
fields, as required by gauge invariance. Moreover, the form of generalized 
diffeomorphism and $\star$-product should be understood.

Finally let us mention that even if we  have restricted our 
analysis to the bosonic sector of the heterotic string,  the inclusion of 
fermions could also be addressed by invoking 
supersymmetry, generalizing the 
discussion in \cite{aamp} (see also \cite{dftfermions,Berman:2013cli}) where 
``will-be massless fermions at a fixed point'', specifically  for modes 
in $\check G_{n_c}$, were considered. From a duality invariant field 
theory point of view,  an uplift including fermions would require an 
analysis from an Extended Field Theory (EFT)\cite{em} in order to include 
magnetic modes. The recent work in in 
Ref.\cite{Blair:2018lbh} might be helpful in this direction.

\newpage

\section*{Acknowledgments}

We thank D. Marqu\'es, E. Malek, C. Nu\~nez and A. Rosabal for useful discussions and the referees for their helpful observations.
This work is partially supported  by CONICET grant PIP-11220110100005  and   
PICT-2016-1358.
G. A. thanks S. Watamura and GGPU program at  Tohoku University for hospitality 
and support.

\appendix

\section{Some Heterotic string basics}
\label{sec:Heterotic string basics}

We summarize here some string theory ingredients (that can be found in string 
books)  needed in the body of the 
article. We mainly concentrate in the $SO(32)$ string.

For a heterotic string compactified to $d$ space-time dimensions, 
Left and Right momenta are encoded in momentum 
\begin{equation}
        {\mathbb L}=(l_L, l_R),
\label{generalizedmomentum}
\end{equation}
defined on a self-dual lattice 
$\Gamma_{26-d,10-d}$ 
of signature $(26-d,10-d)$. By writing  $l_L^{\hat{I}}=(L_L^I,p_{L}^m)$ 
with 
$I=1,\dots,16$ and $m=1,\dots 10-d=r$, the moduli dependent momenta, read
\bea
\label{leftrightmomenta}
L_L^I&=& P^I+RA_{n}^I \tilde p^n\\\nn
l_{L}^{m}&=&{\frac{\sqrt{\ap}}{2}}\big[\frac{\tilde 
p^m}{\tilde R}+2 g^{mn}(\frac{p_{n}}{R}-\frac12B_{nr}
\frac{\tilde p^r}{\tilde R})
-P^I A^{m}_I-
\frac{R}{2}A^{m}_I A_{n}^I\tilde p^n\big]\\\nn
l_{R}^{m}&=&{\frac{\sqrt{\ap}}{2}}\big[-\frac{\tilde 
p^m}{\tilde R}+2g^{mn}(\frac{p_{n}}{R}-\frac12B_{nr}
\frac{\tilde p^r}{\tilde R})
-P^I A^{m}_I-
\frac{R}{2}A^{m}_I A_{n}^I\tilde p^n\big],
\eea
where $g_{mn}, B_{mn}$ are internal metric and antisymmetric tensor 
components,  $A_{m}$ are Wilson lines and $ p_{n}$ and $\tilde p^n$ 
are integers corresponding to KK momenta and windings, respectively. $P_I$ are
$\Spin(32)$ 
weight components.

More schematically, by defining the vector $\check {\mathbb 
P}=( P_I, p_n,\tilde p^n)$ and $\mathbb L=( L_L^I, l_L^m,l_R^m)$ we can write
\begin{equation}
 \mathbb L={\cal R}(\Phi) \check {\mathbb P},
\end{equation}
where  \begin{eqnarray}
{\cal R}=   \left(\begin{matrix} 1 & 0& R A \\
-{\frac{\sqrt{\ap}}{2}}A& \frac{\sqrt{\alpha'}}{R}g^{-1} &   
\frac{\sqrt{\alpha'}}{2\tilde R}(1-g^{-1}B-\frac{1}2A.A\ap) \\
-{\frac{\sqrt{\ap}}{2}}A&\frac{\sqrt{\alpha'}}{R}g^{-1} &   
\frac{\sqrt{\alpha'}}{2\tilde R}(1-g^{-1}B-\frac{1}2A.A\ap) \\
\end{matrix}\right)
     \end{eqnarray}
performs the change of basis. It also rotates the 
coordinates $\check{\mathbb{Y}}=(y^I,y_m,\tilde{y}^m)$ to 
$\mathbb{Y}=(Y^I,y_{L}^m,y_R^m)$.  In particular it 
transforms the $O(16+r,r)$ 
metric $\eta_C$ defined in \eqref{etac} to 
   \begin{equation}
   \label{etadft}
\eta^{{\cal I} {\cal J}} =\begin{pmatrix} 1_{16}& 0& 0 \\
   0 & 0& 1_r\\
   0&1_r& 0
\end{pmatrix}. 
\end{equation}
Notice that ${\cal R}(\Phi)$ encodes the dependence on moduli.

The  mass formulas for string states are (we mainly use the notation in 
\cite{iu})
  \begin{eqnarray}\nn
   \frac{\ap}2 m^{2}_{L}&=& \frac12 l_L^2+( N-1)\\
    \frac{\ap}2 m^{2}_{R}&=& \frac12l_R^2+ \bar N\, 
,
\label{LRstringmasses}
  \end{eqnarray}
 where $  N=N_B, \bar N=\bar N_B+\bar N_F+\bar E_0$ where $N_B$, $\bar N_B$ 
are  the bosonic L and R-oscillator numbers,  $\bar N_F$ is the R fermion 
oscillator number and $\bar
E_0=-\frac12(0) $ for NS (R) sector. The level matching condition is $ 
\frac12 m^{2}_{L}- \frac12 m^{2}_{R}=0$ or, in terms of above notation
\begin{equation}
 \frac12{\mathbb L}^2=\tilde p.p+\frac12P^2=(1-N+\bar N).
\label{LMCwidings}
\end{equation}

In our discussion we  restrict   to  $\bar N_B=0, N_F=\frac12$, 
namely $\bar N=0$. The ``charged vectors'' sector, corresponds to $N=0$, i.e.
${\mathbb L}^2=2$. 
Massless vectors are a particular case with  
$\frac12 l_L^2=1$, $l_R=0$ \footnote{
The normalizations are chosen such that, $\frac{l_L^m}{\sqrt{\ap}}$ , for an 
enhancement point, correspond to the coordinates of the 
weight vectors of a representation in  the lattice span by  simple roots
$\alpha_m$ with $\alpha_m^2=2$.}. 

As is well known, there are $10-d+16$ Left gauge bosons corresponding to 
16 Cartan 
generators $\partial_z Y^I \tilde\psi^{\mu}$ of the original gauge algebra 
as well as $ 10-d$  KK Left gauge bosons 
coming from a Left combination of the metric and antisymmetric field 
$\partial_z Y^m\tilde\psi^{\mu}$. The $10-d$  Right combinations $\partial_z 
X^{\mu}\tilde\psi^{m}$ with 
$m=1,\dots 10-d$ generate the Right Abelian group. These states have $l_R=0$ 
and $l_L=0$, with vanishing winding and KK momenta. 

\section{The $\star$-product}
\label{sec:The star product}
A  $\star$-product, was proposed in \cite{leigh} in 
order to incorporate, in a ``Double Field theory'' description, information 
about bosonic string vertex cocycle factors. If $\check{\mathbb{P}}\equiv 
(p_m,\tilde 
p^{m})$ is an $O(n,n)$ vector encoding information about winding numbers 
$\tilde 
p^{m}$ and Kaluza-Klein (KK) compact  momenta $p_{m}$, then for two fields 
depending on the compact double coordinate $\check{\mathbb Y}\equiv (y^m,\tilde 
y_{m})$ 
their proposed  $\star$-product  reads
\begin{eqnarray}
 \label{eqstarproductFourier}
(\phi_1\star\phi_2)(x,\mathbb{Y})&=&\sum_{\mathbb{P}_1,\mathbb{P}_2}
e^{i{\pi } 
{p}_1 \cdot \tilde p_2 }
\phi^{(\mathbb{P}_1)}_1(x)\phi_2^{(\mathbb{P}_2)}(x)
e^{i(\mathbb{P}_1 + 
\mathbb{P}_2).\mathbb{Y}}\\\nn
&=&\sum_{\mathbb{L}}\left[\sum_{\mathbb{P}_1}
e^{i{\pi } 
{p}_1\cdot (\tilde l-\tilde p_1) }
\phi^{(\mathbb{P}_1)}_1(x)\phi_2^{(\mathbb{L}-\mathbb{P}_1)}(x)\right]e^{
i\mathbb { L
}.\mathbb{Y}}\\\nn
&=& \sum_{\mathbb{L}}(\phi_1\star\phi_2)^{(\mathbb{L})}(x) e^{i\mathbb{L
}.\mathbb{Y}}\\\nn
\end{eqnarray}
where 
\begin{eqnarray}
 \label{starproductmode}
(\phi_1\star\phi_2)^{(\mathbb{L})}(x)&=& 
 \sum_{\mathbb{P}_1}
e^{i{\pi } 
{p}_1 \cdot (\tilde l-\tilde p_1) }
\phi^{(\mathbb{P}_1)}_1(x)\phi_2^{(\mathbb{L}-\mathbb{P}_1)}(x)
\end{eqnarray}
is the Fourier mode of the star product.

It is straightforward to show that the $\star$-product is indeed 
associative.
Namely
\begin{eqnarray}\nn
 \label{assocstarproduct}
\left(\phi_1\star\phi_2)(x,\mathbb{Y}\right)\star\phi_3(x,\mathbb{Y})&=&
\sum_{\mathbb{L}}\sum_{\mathbb{P}_3} e^{i\pi {{l}}\cdot \tilde p_3 }     
(\phi_1\star\phi_2)^{(\mathbb{L})}(x) \phi_3^{(\mathbb{P}_3)}(x)  
e^{i(\mathbb{L}+\mathbb{P}_3).\mathbb{Y}}\\\nn
&=&\sum_{\mathbb{P}_1,\mathbb{P}_2,\mathbb{P}_3}e^{i\pi 
({p_1}+{p_2})\cdot \tilde p_3 }e^{i{\pi } 
{p}_1\cdot \tilde p_2 
}\phi^{(\mathbb{P}_1)}_1(x)\phi_2^{(\mathbb{P}_2)}(x)\phi_3^{(\mathbb{P}_3)}
(x)e^{i(\mathbb{P}_1+\mathbb{P}_2+\mathbb{P}_3).\mathbb{Y}}\\\nn
&=&\sum_{\mathbb{P}_1,\mathbb{P}_2,\mathbb{P}_3}e^{i\pi {p_1}\cdot(\tilde p_2+ 
\tilde p_3) 
}e^{i\pi {p}_2\cdot \tilde p_3} 
\phi^{(\mathbb{P}_1)}_1(x)\phi_2^{(\mathbb{P}_2)}(x)\phi_3^{(\mathbb{P}_3)}
(x)e^{i(\mathbb{P}_1+\mathbb{P}_2+\mathbb{P}_3).\mathbb{Y}}
\\\nn
&=&\sum_{\mathbb{P}_1,\mathbb{L} }e^{i\pi {p_1}\cdot \tilde l }
\phi^{(\mathbb{P}_1)}_1(x)\left[ \sum_{\mathbb{P}_2}
e^{i{\pi } 
{p}_2\cdot (\tilde l-\tilde p_2) 
}\phi^{(\mathbb{P}_2)}_2(x)\phi_3^{(\mathbb{L}-\mathbb{P}_2)}(x)\right]e^{
i(\mathbb{L}+\mathbb{P}_1).\mathbb{Y}}\\
&=&\phi_1(x,\mathbb{Y})\star\left(\phi_2\star\phi_3)(x,\mathbb{Y}\right).
\end{eqnarray}

Interestingly enough, the appearance of  the phases can be traced back 
as a non-commutativity of the string compact coordinates zero modes (see  
\cite{Sakamoto:1989ig}).

 If the sum over  $\phi_i^{(\mathbb{P}_i)}$ modes is constrained by LMC's, 
namely to 
modes satisfying  $\delta (\frac12\mathbb  P_i^{ 2},1-N_i)$ then the same proof 
goes through if we define
\begin{eqnarray}\nn
 \label{starproductmodeconst}
(\phi_1\star\phi_2)^{(\mathbb{L})}(x)&=& 
 \sum_{\mathbb{P}_1}
e^{i{\pi } 
{p}_1\cdot (\tilde l-\tilde p_1) }
\phi^{(\mathbb{P}_1)}_1(x)\phi_2^{(\mathbb{L}-\mathbb{P}_1)}(x)\delta 
(\frac12\mathbb  P_1^{ 2},1-N_1)
\delta (\frac12{(\mathbb{L}-\mathbb{P}_1)}^{ 2},1-N_2).
\end{eqnarray}

Here we just extend this product to account for heterotic string degrees of 
freedom in a $O(r_L,r)$ context. Actually, since it is possible to interpret 
the heterotic string momenta $P^I$ as originating in a 16 dimensional 
torus\cite{Giveon:1994fu}
with some winding and momenta  $(\tilde{p}^I,p_I)$ (with $I=1,\dots 16$) we 
can generalize above expression by including a phase that contains not only 
the compactified winding and momenta but also the gauge ones. More 
concretely, $P_L^I, P_R^I$ can be computed using similar expressions as 
\eqref{leftrightmomenta} above (no Wilson lines) but by imposing $P_R^I=0$. 
Then, 
 $P^I\equiv P_L^I$ root vectors are obtained with,  $G_{IJ}$ the Cartan Weyl 
metric of 
$\Spin(32)$ and $B_{IJ}=G_{IJ}=-B_{JI}$ for $I>J$. It is possible to check then 
that for two vectors $P_1, P_2$ we have $\tilde{p}_1^I 
p_{2I}=\frac12P_1^IE_{IJ}P_{2J}$ 
where 
$E_{IJ}=G_{IJ}+B_{IJ}$.
Therefore, for the heterotic string we would have  (see 
\eqref{generalizedmomentum} above)
$ \mathbb L=(l_L,l_R)\equiv (L_L^I, l_L^{m}, l_R^{m}) $ and  
$\mathbb Y=(y_l,y_R)\equiv (y^I, y_L^{m}, y_R^{m})$ and using that 
\begin{equation}
 {l}_1\cdot \tilde l_2=p_{1m}\tilde{p}_2^{m}+p_{1I}\tilde{p}_2^I 
={p}_1.\tilde p_{2}+\frac12 P_1EP_2,
\label{mwheterotic}
\end{equation}
we recover the expression in \eqref{starheterotic}.
Notice that the phase $\epsilon(P_1,P_2)=e^{i\pi\frac12 P_1EP_2}$ introduces a 
notion of ordering for $\Spin(32)$ roots. For two adjacent roots in the 
corresponding Dynkin diagram  $E_{IJ}=-1$ for $I>J$ and  vanishes otherwise. 
This provides an adequate representation of structure constants for 
$\Spin(32)$ charged operator algebra. Namely 
$[E_{P_1},E_{P_2}]=\epsilon(P_1,P_2)\, E_{P_3}$ (see e.g. the construction in 
\cite{Green:1987sp}).

The same reasoning holds for the full enhanced group. At the enhancement point 
$\Phi_0$  with  $\check {\mathbb 
P}\in \check G_{n_c}(\Phi_0)$, $p_r=0$,  $l_{\hat 
I}$ become the roots of the gauge group and from 
equations \eqref{leftrightmomenta} above we can express windings and momenta in 
terms of the $l_{L}$ such that
\begin{equation}
 {l}_1 \cdot \tilde l_2={p}_{1m} \tilde p_{2}^ m+\frac12 P_1EP_2=l_{1L}{\cal 
E}l_{2L},
\end{equation}
with\cite{Giveon:1994fu}
\begin{equation}
{\cal E}= 
\begin{pmatrix}
(B+g+\frac{\ap}{2} A_IA^I)_{nm} & \sqrt{\ap}g_{nm}A^m_I\\
0  & (G+B)_{IJ}\\
\end{pmatrix}.
\end{equation}
\section{Some useful $SU(3)$ expressions}
\label{sec:su3}

Here we collect some relevant $SU(3)$ conventions used in the example in 
\ref{sec:Gauge symmetry breaking-enhancement along moduli space}. 
 The eight $SU(3)$ generators are denoted by the Cartan generators  $T_1,T_2$ 
and the step raising (lowering) generators $T_\alpha$ ($T_{-\alpha})$ with 
$\alpha\in\{\alpha_1,\alpha_2,\alpha_3\}$, $(\alpha_3=\alpha_1+\alpha_2)$. In particular, they 
must satisfy
\begin{eqnarray}
[ T_i, T_\alpha ] &=& \alpha^m T_\alpha \label{cartancargado} \\ [ 0cm ]
[ T_\alpha, T_{-\alpha} ] &=& \alpha^m T_m  .\label{cargadoconsumenos}
\end{eqnarray}
Here we choose a simple root $\alpha_1,\alpha_2$  basis with $R^2$ coordinates 
\begin{eqnarray}
\label{simplesu3roots}
\alpha_1&=& (\sqrt{2};0),\qquad 
\alpha_2=(-\frac1{\sqrt{2}};\sqrt{\frac32})    \\
\omega_1&=& (\frac1{\sqrt{2}};\frac12\sqrt{\frac23}),\qquad 
\omega_2=(0;\frac12\sqrt{\frac23})
\end{eqnarray}
where $\omega_j$ are the fundamental weights, {\em i.e.}, the
dual basis to the roots: $\omega_j.\alpha^m=\delta_j ^m$.
A weight vector can be expressed in either root or fundamental weight basis as  
$\Lambda=a_1\omega_1+\nobreak a_2\omega_2 =  \lambda_1 \alpha_1 +\lambda_2 \alpha_2$ 
where $a_i$ are the Dynkin labels.
The fundamental $\bf 3$ representation corresponds to 
\begin{center}
$
\begin{array}{cc|cc|cc}
\multicolumn{2}{c|}{\textrm{Dynkin}}  & \lambda_1 & \lambda_2 &
\multicolumn{2}{c}{\textrm{original basis}}  \\
0 & 1 & \frac{1}{3} & 
\frac{2}{3}& 0& \sqrt{\frac{2}{3}}  \\
1 & -1 &  \frac{4}{3} & \frac{2}{3}  & -\frac{1}{\sqrt{2}} & -\frac{1}{\sqrt{6}} 
\\
 -1 & 0 &  -\frac{2}{3} & -\frac{2}{3}&-\frac{1}{\sqrt{2}} & 
-\frac{1}{\sqrt{6}} 
\end{array}
$
\end{center}
The weights for the $\bf 8$ adjoint representation and  $\bf 6$
symmetric representation are given in Tables \ref{table:su3rep8} and  
\ref{table:su3rep6} respectively. 
In the basis of Table \ref{table:su3rep6}, the six dimensional $\bf 6$ 
representation of $SU(3)$ generators 
read

{\tiny{
\begin{eqnarray}
 \label{T6}
&& T_1 = 
\left(
\begin{array}{cccccc}
\sqrt{2} & 0 & 0 & 0 & 0 & 0\\
0 & 0 & 0 & 0 & 0 & 0\\
0 & 0 & -\sqrt{2} & 0 & 0 & 0\\
0 & 0 & 0 & \frac{1}{\sqrt{2}} & 0 & 0\\
0 & 0 & 0 & 0 & -\frac{1}{\sqrt{2}} & 0\\
0 & 0 & 0 & 0 & 0 & 0\\
\end{array}
\right)
\qquad
T_2 = 
\left(
\begin{array}{cccccc}
\sqrt{\frac{2}{3}} & 0 & 0 & 0 & 0 & 0\\
0 & \sqrt{\frac{2}{3}} & 0 & 0 & 0 & 0\\
0 & 0 & \sqrt{\frac{2}{3}} & 0 & 0 & 0\\
0 & 0 & 0 & -\frac{1}{\sqrt{6}} & 0 & 0\\
0 & 0 & 0 & 0 & -\frac{1}{\sqrt{6}} & 0\\
0 & 0 & 0 & 0 & 0 & -2 \sqrt{\frac{2}{3}}\\
\end{array}
\right)\\\nn
&&\\\nn
&&T_{\alpha_1} =
\left(
\begin{array}{cccccc}
0 & \sqrt{2} & 0 & 0 & 0 & 0\\
0 & 0 & \sqrt{2} & 0 & 0 & 0\\
0 & 0 & 0 & 0 & 0 & 0\\
0 & 0 & 0 & 0 & 1 & 0\\
0 & 0 & 0 & 0 & 0 & 0\\
0 & 0 & 0 & 0 & 0 & 0\\
\end{array}
\right)
\qquad
T_{\alpha_2} =
\left(
\begin{array}{cccccc}
0 & 0 & 0 & 0 & 0 & 0\\
0 & 0 & 0 & 1 & 0 & 0\\
0 & 0 & 0 & 0 & \sqrt{2} & 0\\
0 & 0 & 0 & 0 & 0 & 0\\
0 & 0 & 0 & 0 & 0 & \sqrt{2}\\
0 & 0 & 0 & 0 & 0 & 0\\
\end{array}
\right)
\qquad
T_{\alpha_3} =
\left(
\begin{array}{cccccc}
0 & 0 & 0 & \sqrt{2} & 0 & 0\\
0 & 0 & 0 & 0 & 1 & 0\\
0 & 0 & 0 & 0 & 0 & 0\\
0 & 0 & 0 & 0 & 0 & \sqrt{2}\\
0 & 0 & 0 & 0 & 0 & 0\\
0 & 0 & 0 & 0 & 0 & 0\\
\end{array}
\right)
\end{eqnarray}}}
and $T_{-\alpha}={(T_{\alpha})}^t$.

\end{document}